# The Piezoresponse Force Microscopy of surface layers and thin films: effective response and resolution function

Anna N. Morozovska,[*]

V.Lashkaryov Institute of Semiconductor Physics, National Academy of Science of Ukraine,

45, pr. Nauki, 03028 Kiev, Ukraine

Eugene A. Eliseev

Institute for Problems of Materials Science, National Academy of Science of Ukraine,

3, Krjijanovskogo, 03142 Kiev, Ukraine

Sergei V. Kalinin

Materials Science and Technology Division and Center for Nanophase Materials Science, Oak Ridge National Laboratory, Oak Ridge, TN 37831

Signal formation mechanism of Piezoresponse Force Microscopy of piezoelectric surface layers and thin films on stiff and elastically matched substrates is analyzed and thickness dependence of effective piezoelectric response, object transfer function components and Rayleigh two-point resolution are derived. Obtained exact series and simple Pade approximations can be applied for the effective piezoresponse analytical calculations in the case of films capped on different substrates. The effective piezoresponse is thickness dependent for piezoelectric films on substrates with low dielectric permittivity (extrinsic size effect), whereas the thickness dependence is essentially suppressed for giant permittivity or metallic substrates. Thus implications of analysis for ferroelectric data storage and device applications are discussed.

---

[*] Corresponding author, morozo@i.com.ua

## 1. Introduction

Piezoelectric, pyroelectric and dielectric behavior at surfaces or interfaces of functional ferroic materials has attracted much attention due to their importance in device applications. Several groups analyzed the effect of surface layers on polar materials properties and hysteresis loop features within the framework of Landau-Ginzburg-Devonshire phenomenology (see e.g. Refs. 1, 2, 3, 4, 5, 6, 7, 8). Typically, ferroelectric coupling is reduced at the surfaces, giving rise to surface layers with reduced effective Curie temperature. Surface stress affects local dielectric susceptibility and polarization switching by reducing or enhancing Curie temperature, or due to the formation of built-in electric fields, affecting ferroelectric and piezoelectric responses.[9]

In centrosymmetrical materials, symmetry breaking at surfaces and interfaces can give rise to surface piezoelectric coupling even in non-polar materials, and a number of novel electromechanical phenomena, including surface piezoelectricity and flexoelectricity, can exist.[10,11] Even for purely non-polar materials such as graphene or carbon nanotubes, local curvature will result in the redistribution of electron density and the formation of a curvature-dependent electric dipole and hence flexoelectric electromechanical coupling.[12] Similar couplings emerge and are well-known in soft condensed matter systems such as liquid crystals and cellular membranes.[13]

Verification of existing theoretical models, design of functional nanomaterials with predetermined properties, and application in various devices (such as ferroelectric micro and nanocapacitors, sensors and actuators, etc.) necessitate the experimental studies of piezoelectric coupling and ferroelectric properties in surface layers. These considerations necessitate spatially resolved studies of the local polarization switching behavior in thin films, possible in the context of Piezoresponse Force Microscopy (PFM).

The development of Piezoresponse Force Microscopy [14, 15] allowed 2D mapping of domain structures and switching behavior in ferroelectric thin films. Recently, a universal approach for the calculation of electromechanical response in PFM based on decoupled theory by Felten et al.[16] and Scrymgeour and Gopalan[17] has been developed. Previously we have derived analytical expressions for PFM response on semi-infinite materials,[18, 19] obtained analytical expressions for PFM resolution function and domain wall profiles, and developed the theoretical framework for interpretation of PFM spectroscopy data.[20] Despite this progress, little is known on the PFM image formation mechanism in thin films and surface layers, i.e. systems inhomogeneous in $z$-direction. When the film thickness is close to the effective probe size, the image formation is expected to be significantly different from the semi-space case due to the electrostatic field spreading below the film.

In this manuscript, we extend the decoupled approach to derive analytical results for the effective piezoelectric response of thin films on a thick substrate with different elastic and dielectric properties. We evolved the theory of the local piezoresponse *extrinsic* size effect predicted in [21] for thin films capped on the nonpiezoelectric bulk with the same elastic and close dielectric properties. Obtained results can be used for effective piezoresponse calculations in the case of ferroelectric or piezoelectric surface layer capped on the nonpiezoelectric bulk with close elastic properties, e.g. for the system thin epitaxial film/thick substrate with a weak lattice mismatch and arbitrary dielectric permittivity. Furthermore, we extend this analysis for films on rigid substrates, corresponding to e.g. ferroelectric polymers and electromechanically active molecular layers on hard substrates.

## 2. Basic equations

### 2.1 Decoupling approximation in resolution function approach

In decoupling approximation, the electric field in the material is calculated using a rigid electrostatic model (no piezoelectric coupling); the strain or stress field is calculated using constitutive relations for a piezoelectric solid, and the displacement field is evaluated using an appropriate Green's function for an isotropic or anisotropic solid. This approach is used below. It substitutes the rigorous solution for coupled electromechanical problem, available only for transversally-isotropic semi-infinite material.[22]

Below we consider the case when the film (or surface layer) dielectric and piezoelectric properties differ from substrate (or bulk). Assuming that piezoelectric coupling is uniform within the layer, the strain piezoelectric coefficient, $d_{klj}$, depend on vertical coordinate $x_3$ as follows:

$$d_{klj}(x_1, x_2, x_3) = \begin{cases} d_{ijk}^S(x_1, x_2), & 0 \leq x_3 \leq h \\ 0, & h < x_3 < \infty \end{cases} \quad (1)$$

Here $d_{ijk}^S(x_1, x_2)$ is surface layer (film) piezoelectric tensor, $h$ is the film thickness. For a thin ferroelectric film $d_{ijk}^S$ can depend not only on polarization spatial distribution, but also on film thickness (intrinsic size effect).

PFM signal formation mechanism, similar to other linear imaging techniques, can be conveniently analyzed using resolution function theory. The phenomenological resolution function theory for PFM has been introduced by Kalinin *et al.*,[23] and corresponding analytical theory was developed by Morozovska et al.[20] for semi-infinite material. Here, we extend this analysis for the layer case. The surface displacement $u_i(\mathbf{x})$ below the tip (i.e. in the point $x_3 = 0$) determined by the piezoelectric layer is given by the convolution of an *ideal image* $d_{klj}$ given by Eq. (1) with the surface and bulk components of the resolution function:[23]

$$u_i^S(\mathbf{x}) = \int_{-\infty}^{\infty} d\xi_1 \int_{-\infty}^{\infty} d\xi_2 \, W_{ijkl}^f(\xi_1, \xi_2) d_{lkj}^S(x_1 - \xi_1, x_2 - \xi_2) \qquad (2)$$

The piezoelectric film resolution function[20] components $W_{ijkl}^f$ are introduced as

$$W_{ijkl}^f(\xi_1, \xi_2) = \int_0^h d\xi_3 \, c_{kjmn} \frac{\partial G_{im}^f(-\xi_1, -\xi_2, \xi_3)}{\partial \xi_n} E_l(\xi_1, \xi_2, \xi_3) \qquad (3)$$

Coefficients $c_{jlmn}$ are components of the elastic stiffness tensors, $E_k(\mathbf{x}) = -\partial \varphi(\mathbf{x})/\partial x_k$ is the *ac* electric field distribution produced by the probe, and $\varphi$ is the *ac* potential distribution. The surface displacement are described by the elastic Green's function $G_{3j}^f$ at $x_3 = 0$ that depends on Young modulus $Y$ and Poisson ratios $\nu$ of the film (usually $\nu \sim 0.25 - 0.35$) as well as the boundary conditions on the film-substrate interface $x_3 = h$ and a free upper surface (at which the normal stress is absent, $\sigma_{3i}(x_3 = 0) = 0$).

The resolution function components $W_{ijkl}^f$ allow calculation of the piezoresponse signal from domain structures, Fourier image $\tilde{d}_{ijk}^S(\mathbf{q})$ of which exists in usual (e.g. domain stripes, rings etc.) or generalized (infinite plane domain wall) sense. Defining tensorial *object transfer function* (OTF) components, $\tilde{W}_{ijkl}^f$, as the Fourier transform of $W_{ijkl}^f$, Eq.(2) can be rewritten for the Fourier transformation of vertical surface displacement $\tilde{u}_3^S(q_1, q_2)$ as:

$$\tilde{u}_3^S(q_1, q_2) = \tilde{d}_{klj}^S(q_1, q_2) \tilde{W}_{3jkl}^f(-q_1, -q_2). \tag{4a}$$

$$\tilde{W}_{3jkl}^f(q_1, q_2) = \int_{-\infty}^{\infty} dk_1 \int_{-\infty}^{\infty} dk_2 \int_0^h d\xi_3 \tilde{G}_{3m,n}^f(q_1 - k_1, q_2 - k_2, \xi_3) c_{kjmn} \tilde{E}_l(k_1, k_2, \xi_3) \tag{4b}$$

where the Green's function derivatives are $\tilde{G}_{3m,n}^f(k_1, k_2, \xi_3) = ik_n \tilde{G}_{3m}^f(k_1, k_2, \xi_3)$ for $n = 1, 2$ and $\tilde{G}_{3m,n}^f(k_1, k_2, \xi_3) = \partial \tilde{G}_{3m}^f(k_1, k_2, \xi_3)/\partial \xi_3$ for $n = 3$.

Here we consider two limiting cases: (a) the film and substrate elastic properties $\nu$ and $Y$ are matched (i.e. very close) and so the displacement $u_i$ and normal components of stress tensor is continuous at $x_3 = h$; (b) the rigid substrate, i.e. the displacement is zero at the interface $u_i(x_3 = h) = 0$. For these limiting cases the Fourier transform of the Green's function $G_{3j}^f(k_1, k_2, x_3 = 0, \xi_3)$ on coordinates $\{x_1, x_2\}$ has the form:

$$\tilde{G}_{3j}^f(k_1, k_2, \xi_3) = \begin{pmatrix} \tilde{G}_{3j}^s(k_1, k_2, 0, \xi_3) - \tilde{G}_{3j}^s(k_1, k_2, h, \xi_3) \phi_3(kh, \nu) + \\ + i(k_1 \tilde{G}_{1j}^s(k_1, k_2, h, \xi_3) + k_2 \tilde{G}_{2j}^s(k_1, k_2, h, \xi_3)) \phi_\perp(kh, \nu) \end{pmatrix}, \tag{5}$$

For the case (a) of the matched film/substrate elastic properties $\phi_3 = 0$ and $\phi_\perp = 0$, and the Green function $\tilde{G}_{3j}^f$ coincides with the one of the elastically isotropic semi-space $\tilde{G}_{ij}^s(k_1,k_2,x_3,\xi_3)$ given in Appendix A. The functions $\phi_3$ and $\phi_\perp$ for the case (b) of a rigid substrate are

$$\phi_3(k\,h,\nu) = \frac{4(1-\nu)(\exp(-k\,h)(2-2\nu-k\,h)+\exp(k\,h)(2-2\nu+k\,h))}{(3-4\nu)(\exp(-2k\,h)+\exp(2k\,h))+2(1+4(1-\nu)(1-2\nu)+2k^2h^2)}, \quad (6a)$$

$$\phi_\perp(k\,h,\nu) = \frac{4(1-\nu)(\exp(-k\,h)(1-2\nu+k\,h)-\exp(k\,h)(1-2\nu-k\,h))}{k((3-4\nu)(\exp(-2k\,h)+\exp(2k\,h))+2(1+4(1-\nu)(1-2\nu)+2k^2h^2))}, \quad (6b)$$

where $k = \sqrt{k_1^2 + k_2^2}$.

### 2.2. Effective point charge approach for tip representation

Here we derive the response components for a point charge $Q$ located at the distance, $d$, from the surface. For the case of dielectrically transversely isotropic piezoelectric surface layer or film, the electrostatic potential within the layer $\varphi$ created by the point charge $Q$ located outside the layer has the Fourier image:

$$\tilde{\varphi}(k,\xi_3) = \frac{Q}{2\pi\varepsilon_0}\sum_{n=0}^{\infty}\chi^n\left(\frac{\exp\left(-kd-\frac{2hk}{\gamma}n-k\frac{\xi_3}{\gamma}\right)}{k(\varepsilon_e+\kappa)} - \frac{\exp\left(-kd-\frac{2hk}{\gamma}(n+1)+k\frac{\xi_3}{\gamma}\right)}{k(\varepsilon_e+\kappa)(\kappa_b+\kappa)/(\kappa_b-\kappa)}\right) \quad (7)$$

Here the parameter $\chi = \left(\frac{\kappa_b-\kappa}{\kappa_b+\kappa}\right)\left(\frac{\varepsilon_e-\kappa}{\varepsilon_e+\kappa}\right)$ (always $|\chi|\le 1$), $\varepsilon_e$ is the dielectric constant of the ambient, $\kappa = \sqrt{\varepsilon_{33}\varepsilon_{11}}$ is effective dielectric constant, $\gamma = \sqrt{\varepsilon_{33}/\varepsilon_{11}}$ is the dielectric anisotropy factor of the film (or surface layer), $\kappa_b = \sqrt{\varepsilon_{33}^b\varepsilon_{11}^b}$ is effective dielectric constant of the substrate (or bulk); $-d$ is $x_3$-coordinate of the point charge $Q$ (see Fig. 1a). In ultra-dry or inert atmosphere the ambient permittivity $\varepsilon_e = 1$; whereas $\varepsilon_e = 81$ if the water meniscus is formed below the tip. The electric field components are $\tilde{E}_{1,2}(k_1,k_2,\xi_3) = ik_{1,2}\tilde{\varphi}(k,\xi_3)$ and $\tilde{E}_3(k,\xi_3) = -\partial\tilde{\varphi}(k,\xi_3)/\partial\xi_3$ (see Appendix B for details).

Potential distribution calculated for PbTiO$_3$ film on substrates with different dielectric properties is illustrated in Fig. 1 (b,c,d). Note that the electric field becomes more uniform in the vicinity of tip-surface contact ($x_1 = x_2 = x_3 = 0$) with the increase of substrate permittivity $\kappa_b$. Also the increase of the field strength is apparent, since the potential drop across the film thickness $h$ at $x_1 = 0, x_2 = 0$ increases from $0.06U$ (b) to $0.35U$ (c) and $U$ (d) with the increase of $\kappa_b$. The case (d) corresponds to the conducting (e.g. metallic) substrate with $\kappa_b \to \infty$. For the case of semiconducting substrate the conductivity effect (d) could be smeared by the possible appearance of carriers-depleted interface layer with additional potential drop.

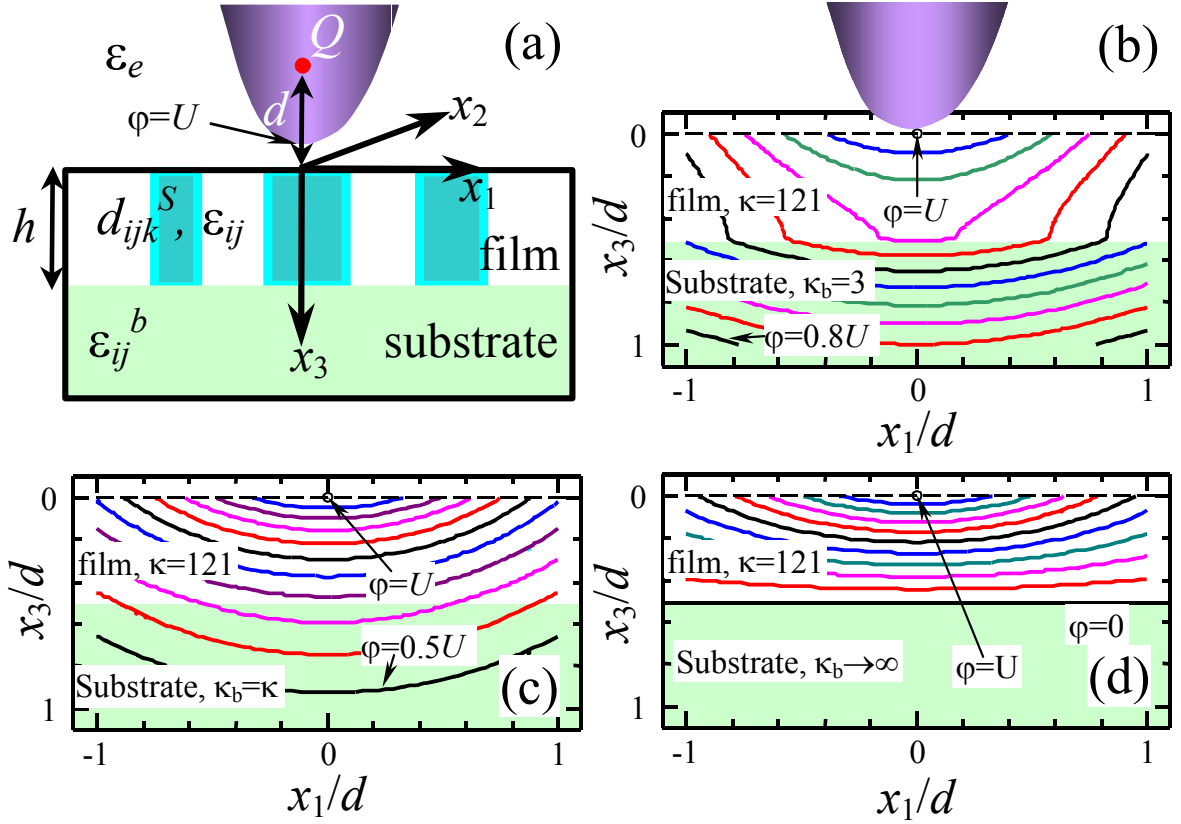

FIG. 1. (Color online). (a) Schematic representation of the considered system. (b, c, d) Equipotential lines of applied electric field inside of PbTiO$_3$ ($\kappa=121$) films with thickness $h=0.5d$ on the substrates with different dielectric permittivity values $\kappa_b = 3;\ 121;\ \infty$ (panels b, c and d respectively); $\varepsilon_e = 1$. The potential difference between the successive lines is fixed to $0.02U, 0.05U, 0.1U$ (panels b, c and d respectively). Lines with maximal and minimal values of potential are marked. Dashed line represents the sample surface $x_3 = 0$.

Since the response Eq. (4) is linear function of the applied electric field, the point charge potential Eq. (7) provides the basic model, and the results for the realistic tip geometries can be obtained using an appropriate image charge model with the help of the additional summation or integration in Eq. (7) over the set of charges representing the tip. For instance a set of point charges $\{d_i, Q_i\}$ can represent monopole (dipole, quadruple, multipole) tips. Conical part of tip can be modeled by the linear charge.

While readily applicable to systems with one boundary (like "ambient-dielectric") the well-known corresponding image-charge expansions for the spherical or planar tips, appropriate image charge series for the considered case of the two boundaries (triple system "ambient-dielectric 1-dielectric 2") are significantly more complex. In general case of multilayer system with several boundaries, rigorous consideration leads to the cumbersome integral equations, making analytical calculations extremely difficult and amenable only to numerical simulations.[24] For these realistic cases, we consider several models, in which we derive the response components as a function of the tip effective charge $Q(h,d)$ located at distance $d(h)$ from the film of thickness, $h$. In this class of models, the parameters of the charge (or series of charges) are chosen so that to reproduce specific features of the external electric field, e.g. location of the isopotential surface representing the conductive tip, tip charge, or radius of curvature. Note that since point charge model is defined by two parameters, only 2 tip parameters can be defined.

Under the contact condition, $\varphi(\mathbf{x}=0)=U$, the ratio $Q/d$ that determines PFM response amplitude universally scales with thickness $h/d$ as (see Appendix B for details):

$$\frac{Q(h,d)}{2\pi\varepsilon_0(\varepsilon_e+\kappa)d} = U \cdot \psi(h,d),$$

$$\psi(h,d) = \left(\sum_{n=0}^{\infty} \chi^n \left(\frac{\gamma d}{\gamma d + 2hn} - \frac{\kappa_b - \kappa}{\kappa_b + \kappa}\frac{\gamma d}{\gamma d + 2h(n+1)}\right)\right)^{-1}. \quad (8)$$

The approximate expression for $\psi(h,d)$ is $\psi^{-1}(h,d) \approx 1 + \left(\frac{\kappa_b + \varepsilon_e}{\kappa - \kappa_b} + \frac{h}{\gamma d}\frac{\varepsilon_e - \kappa}{\kappa \ln(1-\chi)}\right)^{-1}$.

Effective point charge approach in certain cases can be used to describe the spherical or flattened tip in the immediate vicinity or contact of sample surface [25], corresponding to PFM experiments typical geometry (see Table 1).

Table 1. Effective charge approach for different tip models in comparison with capacitance approximation.

| Tip Model | Effective charge parameters |
|---|---|
| **Sphere-plane model**<br><br>Reproduces the electric field of conductive spherical tip with curvature $R_0$, located near the surface at distance $\Delta R \ll R_0$. | Overall charge $Q(h) = C_t^{sph}(h)U$. At $\Delta R \ll R_0$ tip capacitance is:<br><br>$$C_t^{sph}(h) = \begin{cases} 4\pi\varepsilon_0\varepsilon_e R_0 \dfrac{\kappa+\varepsilon_e}{\kappa-\varepsilon_e}\ln\left(\dfrac{\varepsilon_e+\kappa}{2\varepsilon_e}\right)\left(1+\dfrac{\gamma R_0}{h}\dfrac{2\varepsilon_e\kappa\ln(1-\chi)}{(\varepsilon_e-\kappa)^2}\ln\left(\dfrac{\varepsilon_e+\kappa}{2\varepsilon_e}\right)\right), & h \gg R_0 \\ 4\pi\varepsilon_0\varepsilon_e R_0 \dfrac{\kappa_b+\varepsilon_e}{\kappa_b-\varepsilon_e}\ln\left(\dfrac{\varepsilon_e+\kappa_b}{2\varepsilon_e}\right), & h \to 0 \end{cases}$$<br><br>Effective distance $d(h) \approx \dfrac{C_t^{sph}(h)}{2\pi\varepsilon_0(\varepsilon_e+\kappa)}\left(1+\left(\dfrac{\kappa_b+\varepsilon_e}{\kappa-\kappa_b}+\dfrac{h}{\gamma R_0}\dfrac{\varepsilon_e-\kappa}{\kappa\ln(1-\chi)}\right)^{-1}\right)$.<br><br>The field structure is adequately described in the most part of piezoresponse volume, since $\varphi(0,0,-\Delta R)=U$ and $\varphi(r \gg R_0) \sim Q/r$. For $R_0 = 10-100$ nm, $\Delta R \leq 0.1-1$ nm, $\kappa = 50-500$ and $h \gg \gamma R_0$: $d = 0.02-2$ nm at $\varepsilon_e = 1$; $d = 15-150$ nm at $\varepsilon_e = 81$. |
| **Effective point charge model**[26]<br><br>Reproduces the electric field of conductive spherical tip with curvature $R_0$ immediately below the tip apex. | Isopotential surface $\varphi(\mathbf{x}) = U$ has the tip curvature $R_0$ in the point $\{0,0,x_3 = -\Delta R\}$. At $\Delta R \ll R_0$ Pade approximations for effective distance and charge are valid:<br><br>$d(h) \approx \varepsilon_e R_0 \dfrac{\kappa_b h^2 + \kappa\gamma^2 R_0^2}{\kappa_b\kappa(\gamma^2 R_0^2 + h^2)}$, $\quad Q(h,d) \approx \dfrac{2\pi\varepsilon_0(\varepsilon_e+\kappa)U d}{1+\left(\dfrac{\kappa_b+\varepsilon_e}{\kappa-\kappa_b}+\dfrac{h}{\gamma d}\dfrac{\varepsilon_e-\kappa}{\kappa\ln(1-\chi)}\right)^{-1}}$<br><br>The field structure is adequately described in the immediate vicinity of the tip-surface junction. For parameters $R_0 = 10-100$ nm, $\Delta R \leq 0.1-1$ nm, $\kappa = 50-500$ and $h \gg \gamma R_0$: $d = 0.2-10$ nm at $\varepsilon_e = 1$; $d = 11-110$ nm at $\varepsilon_e = 81$. |
| **Disk–plane model**<br><br>Reproduces the electric field of | Overall charge $Q(h) = C_t^{disk}(h)U$. In contact ($\Delta R = 0$) disk capacitance is |
| conductive flattened apex represented by the disk of radius $R_0$ that touches the surface. | $$C_t^{disk}(h) = \begin{cases} 4\varepsilon_0(\varepsilon_e + \kappa)R_0\left(1 - \dfrac{2\gamma R_0}{\pi h}\dfrac{\kappa \ln(1-\chi)}{\varepsilon_e - \kappa}\right), & h \gg R_0 \\ 4\varepsilon_0(\varepsilon_e + \kappa_b)R_0, & h \to 0 \end{cases}$$ Pade approximations for disk capacitance and effective distance are: $$C_t^{disk}(h) \approx \frac{4\varepsilon_0(\varepsilon_e + \kappa)R_0}{1 + \left(\dfrac{\kappa_b + \varepsilon_e}{\kappa - \kappa_b} + \dfrac{\pi h}{2\gamma R_0}\dfrac{\varepsilon_e - \kappa}{\kappa \ln(1-\chi)}\right)^{-1}}, \quad d(h) \approx \frac{2}{\pi}R_0.$$ The field structure is adequately described in the most part of piezoresponse volume, since $\varphi(\rho \leq R_0, 0) \approx U$ and $\varphi(r \gg R_0) \sim Q/r$. Estimation $d = 6 - 60$ nm for $R_0 = 10 - 100$ nm |
| **Capacitance model** Reproduces electric field far from the tip apex ($\Delta R$ is the tip-surface separation) | For the case $\Delta R \ll R_0$ the tip charge $Q(h) = C_t(h) U$, where $C_t^{disk}(h)$ and $C_t^{sph}(h)$ are listed above. The potential $\varphi(0,0,x_3 = -\Delta R) \neq U$ is not fixed. Used for electric field description far from the tip apex at distances $r \gg R_0$: $\varphi(r) \sim Q/r$. Distance $d$ is undetermined. |
| | For the case $\Delta R > R_0$ the tip charge $Q = C_t U$, where $C_t^{sph} \approx 4\pi\varepsilon_0\varepsilon_e R_0$, $C_t^{disk}(h) \approx 8\varepsilon_0\varepsilon_e R_0$ Distance $d \approx R_0 + \Delta R$ for a spherical tip, whereas $d \approx \Delta R$ for a disk. |

## 3. Piezoresponse of the homogeneous surface layers and thin films

*3.1. Piezoelectric surface layers and thin films on matched substrate*

For the case when the layer and substrate (bulk) elastic properties are matched and $d_{ij}^S$ are constant across the layer, the integration of Eq. (4) with Eqs.(5,7) yields analytical results for the vertical surface displacement:

$$u_3^S(h,d) = -\frac{Q(h,d)}{2\pi\varepsilon_0(\varepsilon_e + \kappa)d}\left(w_{333}^m d_{33}^S + w_{313}^m d_{31}^S + w_{351}^m d_{15}^S\right), \qquad (9)$$

where $w_{3jk}^m$ are the normalized OTF function components $\widetilde{W}_{3jkl}^f(q=0)$, defined as:

$$w_{333}^m(h,d) = \sum_{n=0}^{\infty} \chi^n \left(\frac{\gamma d}{\gamma d + 2hn} + \frac{\kappa_b - \kappa}{\kappa_b + \kappa}\frac{\gamma d}{\gamma d + 2h(n+1)}\right)\frac{h(\gamma d + (1+2n+2\gamma)h)}{(\gamma d + (1+2n+\gamma)h)^2} \qquad (10a)$$

$$w_{351}^m(h,d) = \sum_{n=0}^{\infty} \chi^n \left( \frac{\gamma d}{\gamma d + 2hn} - \frac{\kappa_b - \kappa}{\kappa_b + \kappa} \frac{\gamma d}{\gamma d + 2h(n+1)} \right) \frac{\gamma^2 h^2}{(\gamma d + (1+2n+\gamma)h)^2} \quad (10b)$$

$$w_{313}^m(h,d) = (1+2\nu)w_{313}^v(h) - w_{313}^0(h),$$

$$w_{313}^v(h) = \sum_{n=0}^{\infty} \chi^n \left( \frac{\gamma d}{\gamma d + 2hn} + \frac{\kappa_b - \kappa}{\kappa_b + \kappa} \frac{\gamma d}{\gamma d + 2h(n+1)} \right) \frac{h}{\gamma d + (1+2n+\gamma)h}, \quad (10c)$$

$$w_{313}^0(h) = \sum_{n=0}^{\infty} \chi^n \left( \frac{\gamma d}{\gamma d + 2hn} + \frac{\kappa_b - \kappa}{\kappa_b + \kappa} \frac{\gamma d}{\gamma d + 2h(n+1)} \right) \frac{\gamma h^2}{(\gamma d + (1+2n+\gamma)h)^2}.$$

Under the condition $h/\gamma d > 0.1$, the accuracy about 1-2% can be achieved when using the first four terms $n = 0,1$ (point charge + first three images) in a wide range of $|\chi|$. Under the condition $|\chi| < 0.5$, the first two terms $n = 0$ (point charge + its first image) in Eq.(10) provide the accuracy about 5-10%. Under the condition $h/\gamma d \ll 1$, valid for ultrathin films, the approximate summation over all image charges is required. In this case, components $w_{3jk}^m$ are derived as:

$$w_{333}^m \approx \begin{cases} \dfrac{\kappa_b(\varepsilon_e + \kappa)}{\kappa(\varepsilon_e + \kappa_b)} \dfrac{h}{\gamma d}, & h/\gamma d \ll 1, \\ \left(1 + \dfrac{\kappa_b - \kappa}{\kappa_b + \kappa} \dfrac{\gamma d}{\gamma d + 2h}\right) \dfrac{h(h + (d+2h)\gamma)}{(\gamma d + (1+\gamma)h)^2}, & \begin{array}{l} h/\gamma d \geq 1, \\ |\chi| < 0.5 \end{array} \end{cases} \quad (11a)$$

$$w_{351}^m \approx \begin{cases} \dfrac{\varepsilon_e + \kappa}{\varepsilon_e + \kappa_b} \dfrac{h^2}{d^2}, & h/\gamma d \ll 1, \\ \left(1 - \dfrac{\kappa_b - \kappa}{\kappa_b + \kappa} \dfrac{\gamma d}{\gamma d + 2h}\right) \dfrac{\gamma^2 h^2}{(\gamma d + (1+\gamma)h)^2}, & \begin{array}{l} h/\gamma d \geq 1, \\ |\chi| < 0.5 \end{array} \end{cases} \quad (11b)$$

$$w_{313}^m \approx \begin{cases} (1+2\nu) \dfrac{\kappa_b(\varepsilon_e + \kappa)}{\kappa(\varepsilon_e + \kappa_b)} \dfrac{h}{\gamma d}, & h/\gamma d \ll 1, \\ \left(1 + \dfrac{\kappa_b - \kappa}{\kappa_b + \kappa} \dfrac{\gamma d}{\gamma d + 2h}\right) \left( \dfrac{(1+2\nu)h}{\gamma d + (1+\gamma)h} - \dfrac{\gamma h^2}{(\gamma d + (1+\gamma)h)^2} \right), & \begin{array}{l} h/\gamma d \geq 1, \\ |\chi| < 0.5 \end{array} \end{cases} \quad (11c)$$

Under the condition $h/\gamma d \gg 1$, valid for thick film, Eqs.(9-11) tend to the expression for $u_3$ obtained earlier in Refs. 18, 19 for semi-infinite system, as expected.

Within the framework of the effective point charge approach (see Eq.(8)) the effective vertical piezoresponse $d_{33}^{eff} = u_3(\mathbf{x}=0)/U$ has the form:

$$d_{33}^{eff}(h,d) = -\psi(h,d)\left(w_{333}^m d_{33}^S + w_{313}^m d_{31}^S + w_{351}^m d_{15}^S\right), \tag{12}$$

Exact series for the function $\psi(h,d)$ is given by Eq.(8) in the form of exact series or as Pade approximation. Typical dependence of the normalized response components $w_{3jk} = \psi(h,d) w_{3jk}^m$ on the layer thickness $h/d$ is shown in Fig. 2.

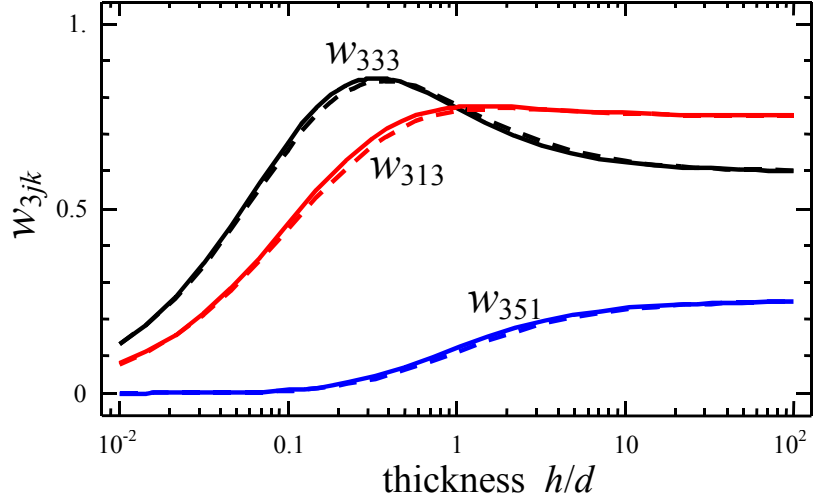

FIG. 2. (Color online). Response components $w_{3jk}$ vs. the layer thickness $h/d$ for Poisson ratio $\nu = 0.3$, dielectric anisotropy $\gamma = 1$ and permittivity $\kappa = 30$ capped on the non-piezoelectric bulk with $\gamma = 1$, $\kappa_b = 260$ (e.g. SrTiO$_3$), the same elastic properties and ambient dielectric constant $\varepsilon_e = 1$. Solid curves correspond to exact series given by Eqs. (8, 10), dashed ones correspond to the first terms with $n = 0$ and Pade approximation for $\psi(h,d)$.

The results in this section can be used to estimate effective electromechanical response for surface piezoelectric layers. Assuming that inversion symmetry breaking in the vicinity of the surface leads to the appearance of spontaneous polarization $P_3^S \cong 1\,\mu C/cm^2$, surface piezoelectric effect coefficients $d_{ij}^S$ of incipient ferroelectric perovskite SrTiO$_3$ can be estimated as follows. Using SrTiO$_3$ bulk electrostriction coefficients $Q_{11} = 0.046\,m^4/C^2$, $Q_{12} = -0.014\,m^4/C^2$, $Q_{44} = 0.019\,m^4/C^2$ (recalculated from Ref. 27) we obtained piezoelectric tensor components $d_{33}^S = 2\varepsilon_0\varepsilon_{33}Q_{11}P_3^S = 2.1\,pm/V$, $d_{31}^S = 2\varepsilon_0\varepsilon_{33}Q_{12}P_3^S = -0.62\,pm/V$ and $d_{15}^S = 2\varepsilon_0\varepsilon_{11}Q_{44}P_3^S = 0.44\,pm/V$. The dependences of effective response $d_{33}^{eff}$ via the thickness $h/d$ of piezoelectric surface layer with different permittivity $\kappa$ capped on the SrTiO$_3$ non-piezoelectric bulk is shown in Fig.3.

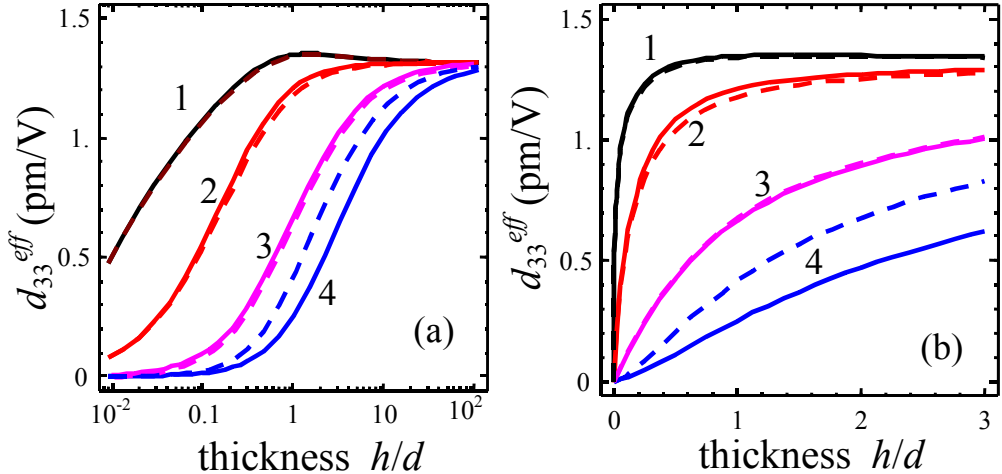

FIG. 3. (Color online). Log-linear plot (a) and linear plot (b) of effective response $d_{33}^{eff}(h)$ dependence on the thickness $h/d$ of SrTiO$_3$ piezoelectric surface layer with Poisson ratio

$\nu = 0.3$, anisotropy $\gamma = 1$ and permittivity $\kappa = 3, 30, 300, 3000$ (curves 1, 2, 3, 4 respectively) capped on the SrTiO$_3$ non-piezoelectric bulk with the same elastic properties and $\kappa_b = 260$, $\varepsilon_e = 1$. Solid curves correspond to exact series given by Eqs.(10); dashed curves are the first terms with $n = 0$ and Pade approximation for $\psi(h,d)$.

It is clear from Figs. 3 and Table 1 that for the layer thickness of 2 nm (5 lattice constants), tip radius $R_0 = 10$ nm response is about 1-1.3pm/V, and for $R_0 = 100$ nm response is about 0.1-1. pm/V depending on $\kappa$ value. Higher measurable response requires sharper tips, i.e. better instrumentation.

It is clear from Figs.2-3, that when the piezoelectric layer dielectric permittivity $\kappa$ lays in region $\varepsilon_e < \kappa < \kappa_b$, the first two terms $n=0$ (point charge + its first image) provide accuracy of better then 5% even in the case $|\chi| \cong 0.95$ (compare dashed and solid curves). In the case $\kappa > \kappa_b$, the series (10) converges more slowly, and accuracy not less than 10% corresponds to the case $|\chi| \leq 0.5$. Note, that the case $\kappa \ll \kappa_b$ may be realized in the systems "high-$\kappa_b$ ferroelectric bulk + low-$\kappa$ piezoelectric surface" or "giant-$\kappa_b$ incipient or close to Curie temperature bulk + piezoelectric surface", whereas the opposite case $\kappa \gg \kappa_b$ corresponds to the systems "high-$\kappa$ piezoelectric surface + non-polar substrate" or "giant-$\kappa$ close to transition temperature surface + paraelectric bulk".

*3.2. Thin piezoelectric film on a rigid substrate*

For the case when the piezoelectric film is epitaxially clamped on a thick rigid substrate, the effective vertical piezoresponse $d_{33}^{eff} = u_3(\mathbf{x}=0)/U$ can be obtained in the form of series similarly to Eqs.(12), using appropriate elastic Green function given by Eqs.(5-6), as:

$$d_{33}^{eff}(h,d) = -\psi(h,d)\left(w_{333}^r d_{33}^S + w_{313}^r d_{31}^S + w_{351}^r d_{15}^S\right), \tag{13}$$

where $w_{3jk}^r(h,d)$ are the normalized components of the OTF function $\widetilde{W}_{3jkl}^f(q=0)$ corresponding to the rigid substrate case. Their exact expressions in the form of series are very cumbersome and could be obtained with the help of results listed in Appendixes A by formalism evolved in Appendix C. Since one-fold integrals are readily available, this problem is beyond the scope of the paper. Here, we derive effective numerical calculations and propose approximate analytical relations instead. After lengthy analysis, the components of $w_{3jk}^r(h,d)$ are derived as

$$w_{333}^r \approx \begin{cases} \dfrac{\kappa_b(\varepsilon_e+\kappa)}{\kappa(\varepsilon_e+\kappa_b)}\left(\dfrac{h}{\gamma d}+\dfrac{(1-2\nu)}{(1-\nu)}\dfrac{h^3}{\gamma d^3}\right), & h/\gamma d \ll 1 \\[2mm] \left(1+\dfrac{\kappa_b-\kappa}{\kappa_b+\kappa}\dfrac{\gamma d}{\gamma d+2h}\right)\dfrac{h(h+(d+2h)\gamma)}{(\gamma d+(1+\gamma)h)^2}, & \begin{array}{l}h/\gamma d \geq 1,\\ |\chi|<0.5\end{array}\end{cases} \tag{14a}$$

$$w_{351}^r \approx \begin{cases} \dfrac{\varepsilon_e+\kappa}{\varepsilon_e+\kappa_b}\left(1-\dfrac{(1-2\nu)}{(1-\nu)}\right)\dfrac{h^2}{d^2}, & h/\gamma d \ll 1 \\[2mm] \left(1-\dfrac{\kappa_b-\kappa}{\kappa_b+\kappa}\dfrac{\gamma d}{\gamma d+2h}\right)\dfrac{\gamma^2 h^2}{(\gamma d+(1+\gamma)h)^2}-\dfrac{1-2\nu}{(1-\nu)}\times \\[2mm] \times\left(\dfrac{\gamma d}{\gamma d+2h\gamma}-\dfrac{\kappa_b-\kappa}{\kappa_b+\kappa}\dfrac{\gamma d}{\gamma d+2h(1+\gamma)}\right)\dfrac{\gamma^2 h^2}{(\gamma d+(1+3\gamma)h)^2}, & \begin{array}{l}h/\gamma d \geq 1,\\ |\chi|<0.5\end{array}\end{cases} \tag{14b}$$

$$w_{313}^r \approx \begin{cases} \dfrac{\kappa_b(\varepsilon_e + \kappa)}{\kappa(\varepsilon_e + \kappa_b)}\left((1+2\nu) - \dfrac{(1-2\nu)(1+\nu)}{(1-\nu)}\right)\dfrac{h}{\gamma d}, & h \ll d \\[2ex] \begin{aligned}&\left(1 + \dfrac{\kappa_b - \kappa}{\kappa_b + \kappa}\dfrac{\gamma d}{\gamma d + 2h}\right)\left(\dfrac{(1+2\nu)h}{(\gamma d + (1+\gamma)h)} - \dfrac{\gamma h^2}{(\gamma d + (1+\gamma)h)^2}\right) - \\ &-\dfrac{(1+\nu)(1-2\nu)}{(1+2\nu)(1-\nu)}\left(\dfrac{\gamma d}{\gamma d + 2h\gamma} + \dfrac{\kappa_b - \kappa}{\kappa_b + \kappa}\dfrac{\gamma d}{\gamma d + 2h(1+\gamma)}\right) \times \\ &\times \left(\dfrac{h(1+2\nu)}{\gamma d + (1+3\gamma)h} + \dfrac{\gamma h^2}{(\gamma d + (1+3\gamma)h)^2}\right)\end{aligned}, & \begin{array}{l} h/\gamma d \geq 1, \\ |\chi| < 0.5 \end{array} \end{cases}$$ (14c)

Under the condition $h \gg d$, the components $w_{3jk}^r \to w_{3jk}^m$, as expected. Below we compare the piezoresponse of films on matched and rigid substrates.

### 3.3. Piezoelectric response of thin films on different substrates

The case of matched substrate could approximate the structures like BaTiO$_3$ or PbTiO$_3$ film on SrTiO$_3$, SrRuO$_3$ or TiO$_2$ substrate with $\kappa_b \approx 100 - 300$. The typical rigid dielectric substrates are MgO oxide, sapphire Al$_2$O$_3$ or carbon with $\kappa_b \approx 5-10$. Silicon ($\kappa_b \approx 3-12$) and SiO$_2$ ($\kappa_b \approx 5$) have smaller elastic stiffness than typical perovskites.

High $\kappa_b$ values correspond to metallic ($\kappa_b \to \infty$) or giant permittivity (e.g. relaxor with $\kappa_b \sim 10^4 - 10^5$) substrates. For a particular case of thin layer on substrate with $\kappa_b \gg \kappa$, one obtains from Eq.(8) that $\psi(h,d) \approx \gamma d \kappa / h(\varepsilon_e + \kappa)$. At $\kappa_b/\kappa \to \infty$ the contributions $w_{333} = \psi(h,d)w_{333}^f$ and $w_{313} = \psi(h,d)w_{313}^f$ ($f = m$ or $f = r$) tend to constant values at $h/\gamma d \ll 1$, whereas $w_{351} = \psi(h,d)w_{351}^f$ tends to zero allowing for Eqs.(12, 14).

Size effect on effective piezoelectric response $d_{33}^{eff}$ and its components $w_{3jk} = \psi(h,d)w_{3jk}^{f}$ of the thin films is illustrated in Figs. 4-6 for BaTiO$_3$, PbTiO$_3$ and LiNbO$_3$ thin films capped on rigid and matched substrates with different dielectric constants $\kappa_b$.

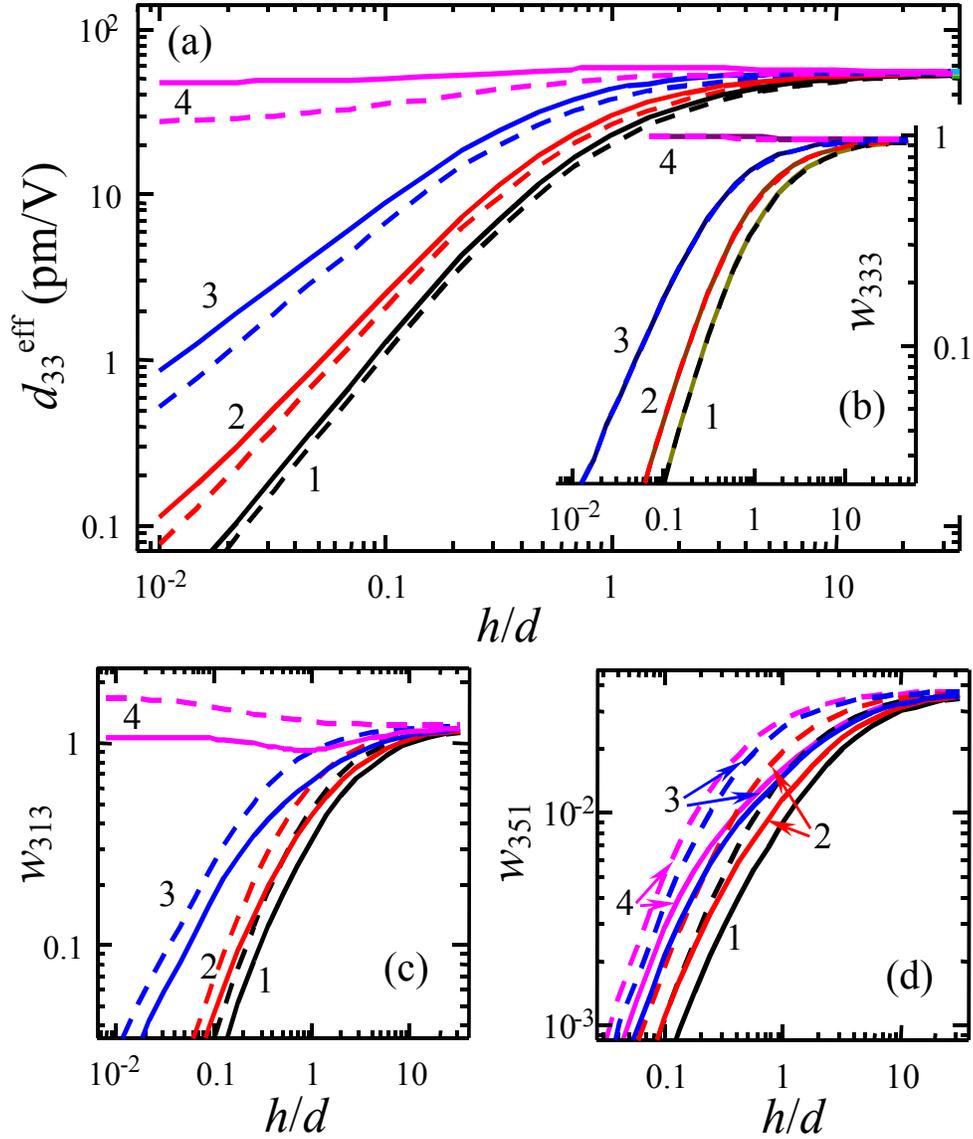

FIG 4. (Color online). Dependence of vertical PFM response $d_{33}^{eff}$ (a) and its components $w_{3jk}$ (b,c,d) on film thickness $h$ for BaTiO$_3$ ($\nu = 0.35$, $\kappa = 700$, $\gamma = 0.24$, $d_{33}^{S} = 85.6$, $d_{15}^{S} = 392$,

$d_{31}^S = -34.5$pm/V) films on the non-piezoelectric rigid (solid curves) and matched (dashed curves) substrate with different dielectric constants $\kappa_b = 3; 30; 300; \infty$ (curves 1, 2, 3 and 4 respectively); $\varepsilon_e = 1$.

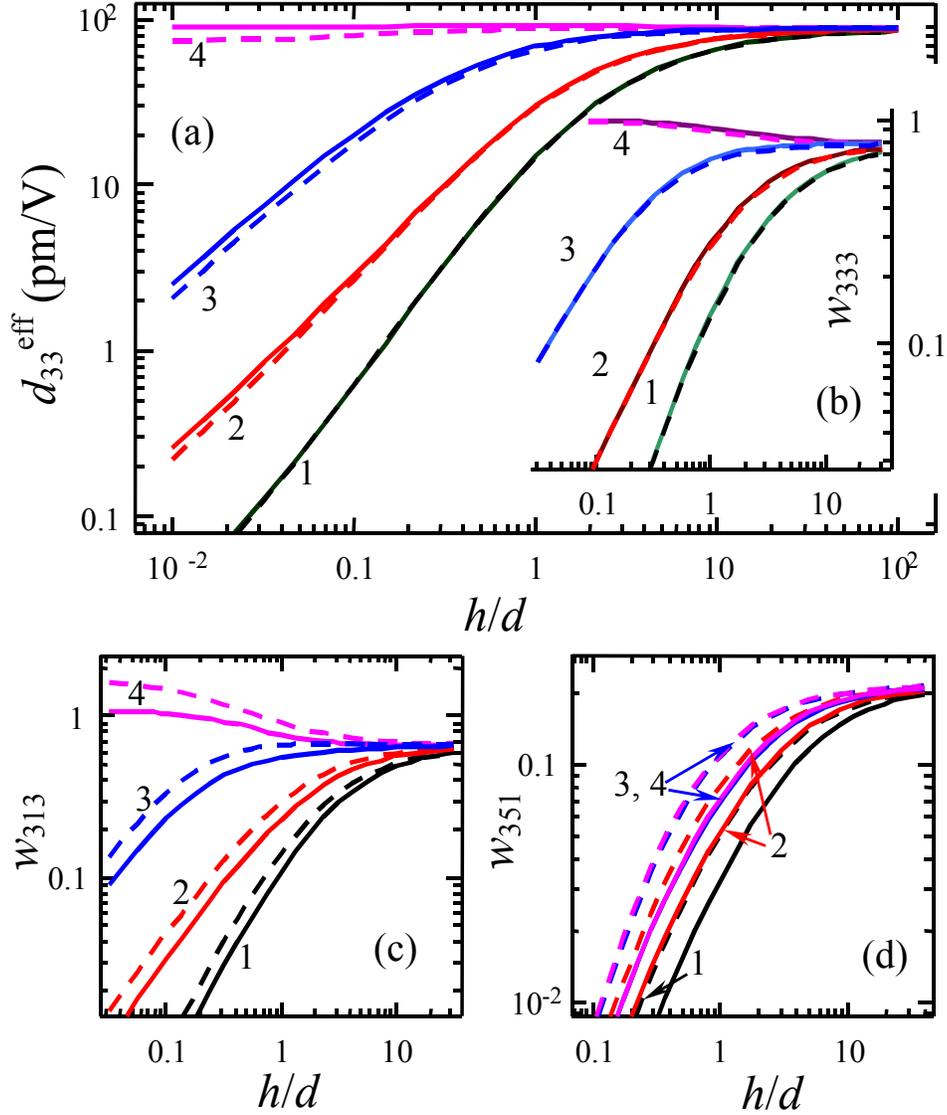

FIG 5. (Color online). Dependence of vertical PFM response $d_{33}^{eff}$ (a) and its components $w_{3jk}$ (b,c,d) on film thickness $h$ for PbTiO$_3$ ($\nu = 0.35$, $\kappa = 121$, $\gamma = 0.87$, $d_{33}^S = 117$, $d_{15}^S = 61$, $d_{31}^S = -25$pm/V) films on the non-piezoelectric rigid (solid curves) and matched (dashed

curves) substrate with different dielectric constants $\kappa_b = 3; 30; 300; \infty$ (curves 1, 2, 3, 4 respectively); $\varepsilon_e = 1$.

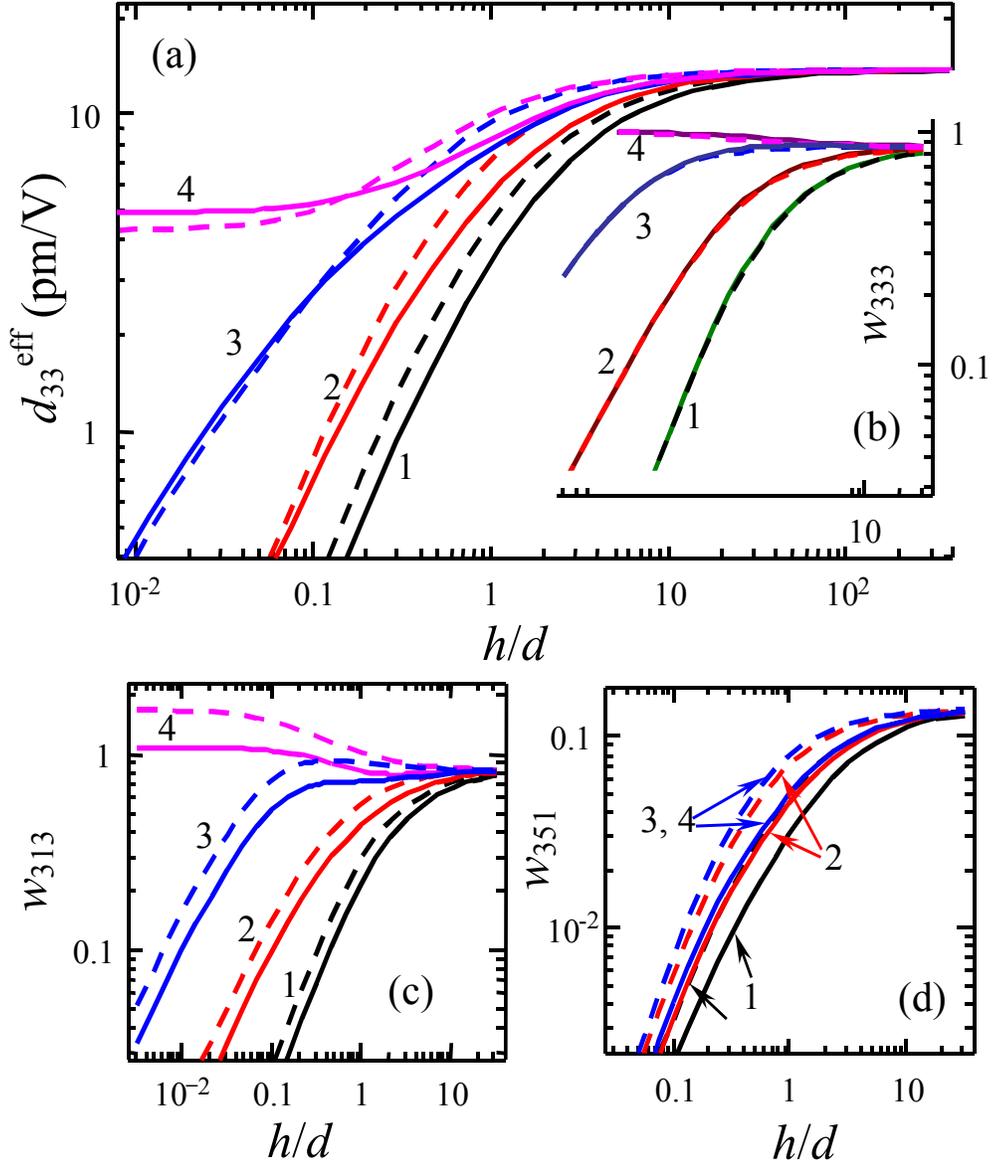

FIG 6. (Color online). Dependence of vertical PFM response $d_{33}^{eff}$ (a) and its components $w_{3jk}$ (b,c,d) on film thickness $h$ for LiNbO$_3$ ($\nu = 0.35$, $\kappa = 50$, $\gamma = 0.60$, $d_{33}^S = 6$, $d_{15}^S = 68$, $d_{31}^S = -$1pm/V) films on the non-piezoelectric rigid (solid curves) and matched substrate (dashed

curves) with different dielectric constants $\kappa_b = 3;\ 30;\ 300;\ \infty$ (curves 1, 2, 3, 4 respectively); $\varepsilon_e = 1$. Note, that $d_{22}^S$ does not contribute into the homogeneous layer response.[20]

It is clear from a comparison between curve 4 and curves 1,2,3 in Figs.4-6, that for piezoelectric films on high-$\kappa_b$ substrate thickness dependence of PFM response is essentially suppressed ($d_{33}^{eff}(h) \approx const$). This fact can be explained by the thickness-induced changes of the external field structure inside the film: both the strength and homogeneity of the field increases with substrate permittivity increase (see Figs. 1 (b,c,d) and comments to them). Under the absence of intrinsic size effects the piezoelectric response in the homogeneous electric field is independent on film thickness (see, e.g. Appendix A in Ref. 28).

It is follows from all the Figs. 4-6 that $d_{33}$ contribution $w_{333}$ differs very little (<5%) for the case of rigid substrate in comparison with the case matched one in the wide range of dielectric constants $\kappa$, $\kappa_b$ and anisotropy $\gamma$. This non-trivial result becomes clear from comparison of Eq.(14a) and (11a). From Eq.(11a) the longitudinal OTF component scales as $w_{333}^m \sim h$ at small thickness $h$, whereas the second term in Eq.(14a) is the difference $\left(w_{333}^m - w_{333}^r\right) \sim (h/d)^3$, which is negligibly small at small $h$ values. Under the condition $h > d$ we obtained that $w_{333}^m \approx w_{333}^r$ from rigorous series expansions.

The piezoresponse contributions $w_{313}$ and $w_{351}$ (coupled with $d_{31}$ and $d_{15}$ respectively) depend strongly on substrate stiffness (up to several times). These thickness variations are more pronounced for $0.1 < h/d < 10$ and disappear at $h/d > 10^2$. The result is clear after the comparison of Eqs.(14b,c) with (11b,c). At small thickness the differences

$\left(w_{351}^m - w_{351}^r\right) \sim (h/d)^2$ and $\left(w_{313}^m - w_{313}^r\right) \sim h/d$ are the same order as the values $w_{351}^m$ and $w_{313}^m$ correspondingly.

The above consideration explains the fact that the difference between the effective piezoresponse for a PbTiO$_3$ film on matched and rigid substrate are the smallest in comparison with BaTiO$_3$ and LiNbO$_3$ films, since the relative contribution of $d_{33}^S$ is the highest for PbTiO$_3$. The increase of PFM response for the case of PbTiO$_3$ and BaTiO$_3$ films on a rigid substrate in comparison with the ones on elastically matched substrates can be attributed to the negative value of $d_{31}$, which contribution decreases for the case of rigid substrate, whereas the contribution of $d_{15}$ is negligible for these materials.

The difference in piezoresponse should be more pronounced for the case of a film on a rigid substrate in comparison with a freestanding film. The case of a freestanding film could be considered within the framework of proposed approach, keeping in mind that the appropriate Green function should satisfy the condition $\sigma_{3i} = 0$ on both surfaces $x_3 = 0, h$. The freestanding BaTiO$_3$ thin film dielectric response was studied by Scott et al [29]. To the best of our knowledge the PFM response measurements of a freestanding film are absent.

Here, the effective piezoresponse appears thickness dependent due to the finiteness of the signal generation volume and the thickness dependent structure of electrostatic potential. The effect should be clearly distinguished from the intrinsic size effects related to the inhomogeneous polarization distribution in finite structures (see e.g. Refs. 1, 3) that in beyond the scope of the paper. Usually $d \sim 1 - 100$ nm, thus piezoelectric response appeared thickness-dependent for the films thickness less than 100nm.

In realistic experiments predicted extrinsic size effect would interfere with several intrinsic ones caused by film-substrate misfit strain, correlation volume decrease and depolarization field [8, 30, 31]. For thin films the intrinsic effects lead to the dielectric permittivity $\varepsilon_{ii}(h)$ and spontaneous polarization $P_S(h)$ thickness dependence. Keeping in mind that piezoelectric constants $d_{ij}^S \sim \varepsilon_{ij} P_3^S$, they are dependent on the thickness $h$, the dependence $d_{ij}^S(h)$ should be included in Eqs.(9, 12, 13).

Besides renormalization,[30] the size-driven phase transition into paraelectric phase appeared in the thin films at thickness $h = h_{cr}$, where the critical thickness $h_{cr}$ depends on temperature, stress *etc*. In accordance with recent experimental data [Ref. 32, 33, 34] $h_{cr} = 1-2$ nm for PbTiO$_3$ on SrTiO$_3$ substrate and $h_{cr} \leq 5$ nm for BaTiO$_3$ on SrRuO$_3$ substrate at room temperature.

Depending on the ambient dielectric permittivity $\varepsilon_e$ and tip curvature $R_0$, the effective distance $d \sim 1-100$ nm for PbTiO$_3$ and BaTiO$_3$, the considered extrinsic size effect is the most pronounced at $h/d < 10$ (see Figs.4-6), while intrinsic one is significant at thicknesses $h/h_{cr} < 10$, where $h_{cr} \sim 1-5$ nm. Thus, when the values $h_{cr}$ and $d$ are of the same order extrinsic and intrinsic size effects will interfere, however their contributions could be separated for the homogeneous (single-domain) film by appropriate fitting of experimental data by Eq. (12) or (13) with appropriate dependence $d_{ij}^S(h)$. Note that in the analysis above, the piezoelectric coefficients have been chosen equal to the corresponding bulk values.

## 4. Resolution function of thin piezoelectric films

For the case when the considered piezoelectric layer or film is inhomogeneous in the transverse directions $\{x_1, x_2\}$ (e.g. it is divided into polar regions or posses domain structure with different piezoelectric tensors, $d_{ijk}^S(x_1, x_2)$), the Fourier transform of the vertical surface displacement, $\tilde{u}_3^S(\mathbf{q})$, and effective Piezoresponse, $\tilde{d}_{33}^{eff}(\mathbf{q})$, over transverse coordinates are

$$\tilde{u}_3^S(\mathbf{q}) = \tilde{W}_{313}^f(\mathbf{q})\tilde{d}_{31}(\mathbf{q}) + \tilde{W}_{333}^f(\mathbf{q})\tilde{d}_{33}(\mathbf{q}) + \tilde{W}_{351}^f(\mathbf{q})\tilde{d}_{15}(\mathbf{q}), \tag{15a}$$

$$\tilde{d}_{33}^{eff}(\mathbf{q}) = -\psi(h,d)\left(\tilde{w}_{313}^f(\mathbf{q})\tilde{d}_{31}(\mathbf{q}) + \tilde{w}_{333}^f(\mathbf{q})\tilde{d}_{33}(\mathbf{q}) + \tilde{w}_{351}^f(\mathbf{q})\tilde{d}_{15}(\mathbf{q})\right), \tag{15b}$$

Under the condition $|\chi| \leq 0.5$ we derived Pade approximations for the normalized tensorial transfer function components $\tilde{w}_{3ij}^f(\mathbf{q})$ as:

$$\tilde{w}_{313}^f(q,h) \approx \frac{w_{313}^f(h)}{1 + \gamma\, w_{313}^f(h) qd}, \tag{16a}$$

$$\tilde{w}_{333}^f(q,h) \approx \frac{2\, w_{333}^f(h)}{2 + \gamma\, w_{333}^f(h) qd}, \tag{16b}$$

$$\tilde{w}_{351}^f(q,h) \approx \frac{6\, w_{351}^f(h)}{6 + \gamma\, w_{351}^f(h)(qd)^3}, \tag{16c}$$

where $q = \sqrt{q_1^2 + q_2^2}$ and $w_{3jk}^f(h)$ are given by Eqs.(10) or (11) for matched substrate ($f = m$) or by Eqs.(14) for a rigid one ($f = r$). In the limit of thick film ($h/\gamma d \gg 1$) Eqs.(16) converge to the expressions obtained earlier in Ref. 20 for semi-infinite system, as expected (see also Appendix C for details).

Since the piezoelectric tensor components are proportional to the spontaneous polarization vector components, the distribution of $d_{ij}^S$ in the domain structure is determined mainly by the spontaneous polarization direction. In particular, all $d_{ij}^S$ change their sign (but not the absolute value) simultaneously with spontaneous polarization sign change $+P_S \to -P_S$ in 180°-domain structures. Thus, for tetragonal ferroelectrics one can introduce a rotationally invariant transfer function of a vertical PFM response, $F_3(q) \cong \widetilde{W}^f_{313}(q)d^S_{31} + \widetilde{W}^f_{333}(q)d^S_{33} + \widetilde{W}^f_{351}(q)d^S_{15}$, that determines the resolution of 180°-domains.

The spectrum of $F_3(q)$ is shown in Figs. 7-9 for BaTiO$_3$, LiNbO$_3$ and PbTiO$_3$ films of different thickness, $h$, and on the substrates with different dielectric permittivity, $\kappa_b$.

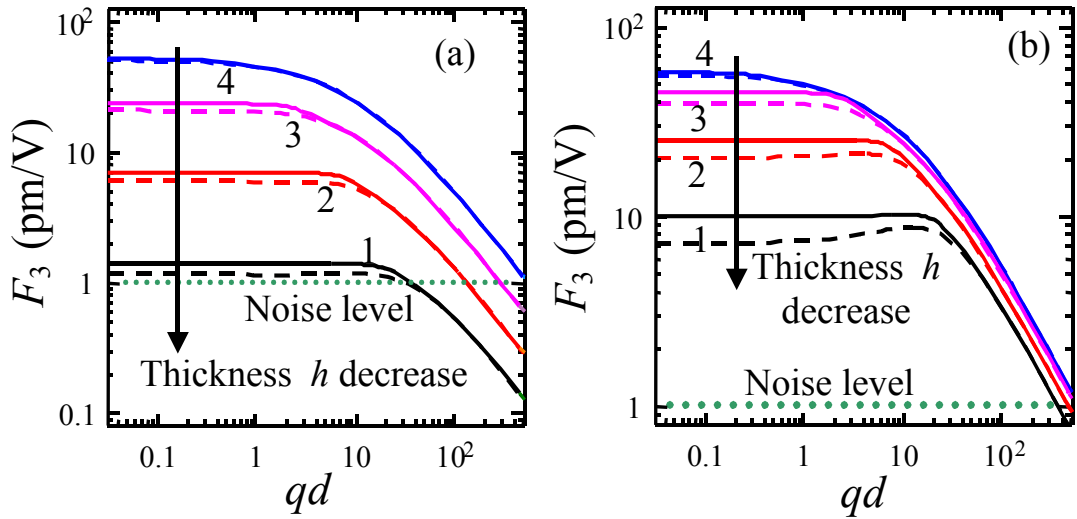

FIG 7. (Color online). Vertical PFM transfer function rotation invariant $F_3(q)$ spectrum for BaTiO$_3$ ($\kappa = 700$) films with thickness $h/d = 0.1, 0.3, 1, 10$ (curves 1, 2, 3, 4) on the non-piezoelectric rigid (solid curves) and matched substrates (dashed curves) with different dielectric constant $\kappa_b = 3; 300$ (panels a, b, respectively).

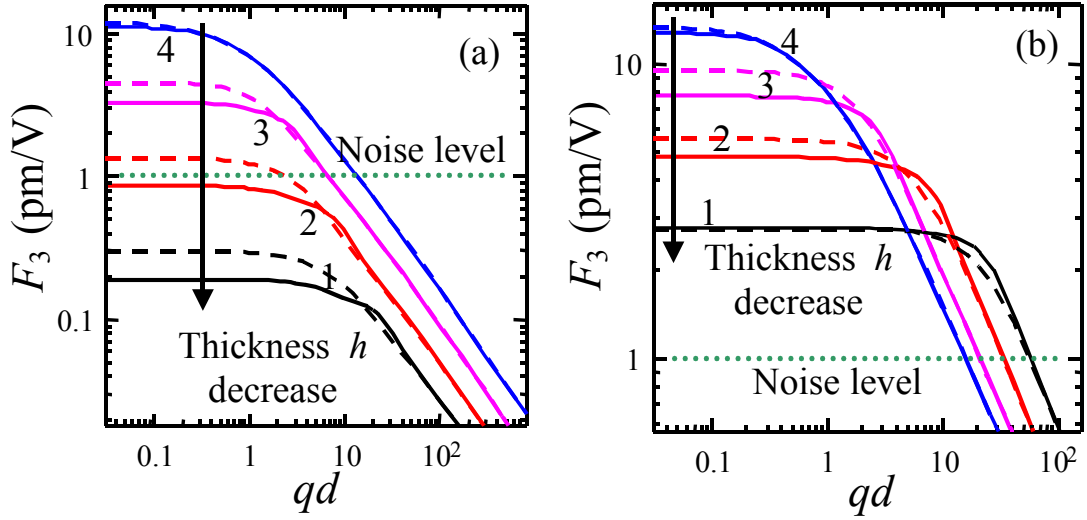

FIG 8. (Color online). Vertical PFM transfer function rotation invariant $F_3(q)$ spectrum for LiNbO$_3$ ($\kappa = 50$) films with thickness $h/d = 0.1, 0.3, 1, 10$ (curves 1, 2, 3, 4) on the non-piezoelectric rigid (solid curves) and matched substrates (dashed curves) with different dielectric constant $\kappa_b = 3; 300$ (panels a, b respectively).

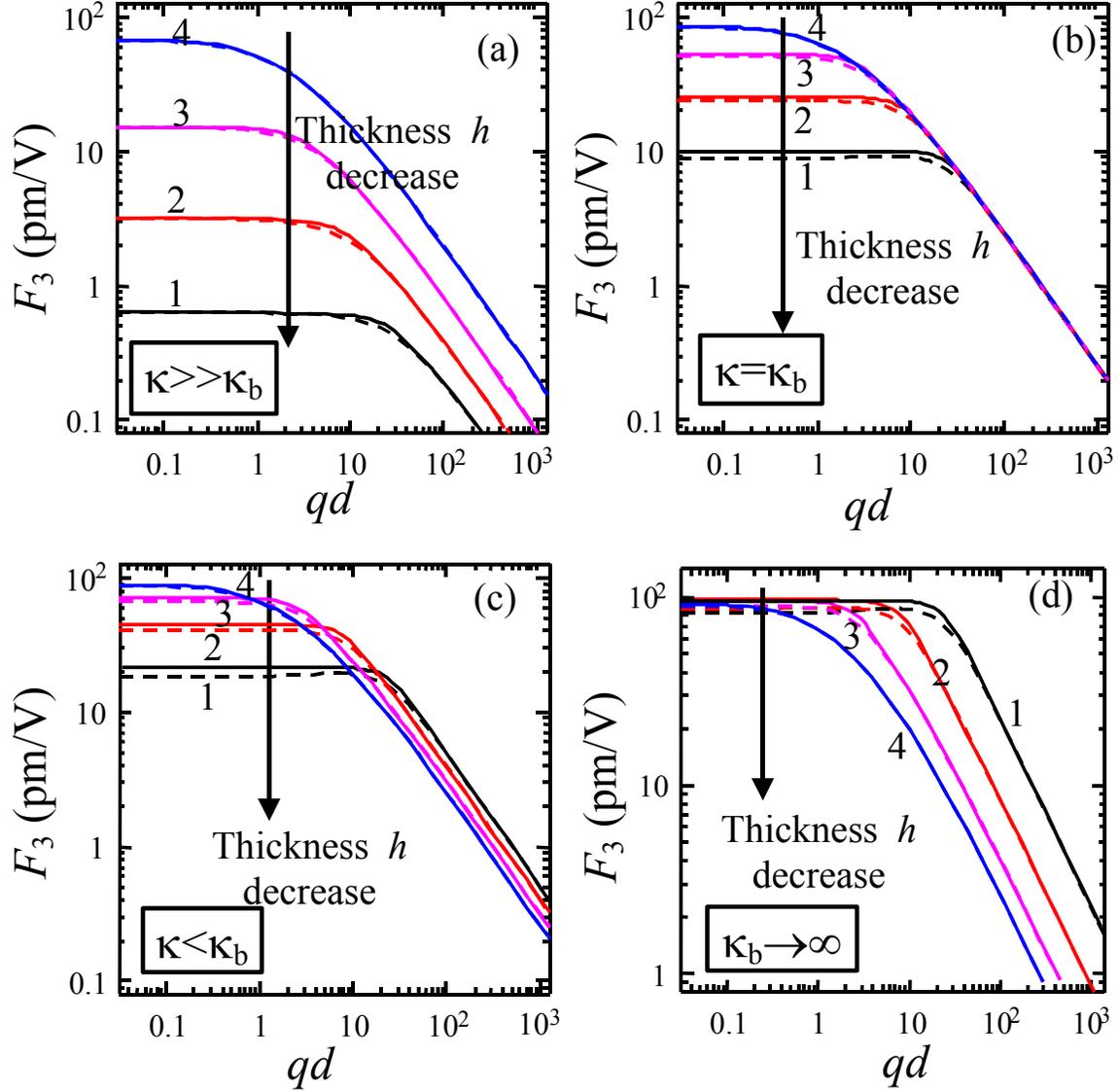

FIG 9. (Color online). Vertical PFM transfer function rotation invariant $F_3(q)$ spectrum for PbTiO$_3$ ($\kappa = 121$) films with thickness $h/d = 0.1, 0.3, 1, 10$ (curves 1, 2, 3, 4) on the non-piezoelectric rigid (solid curves) and matched substrates (dashed curves) with different dielectric constant $\kappa_b = 3; 121; 300; \infty$ (a, b, c, d respectively).

It is clear that the transfer function amplitude decreases with thickness decrease. However the decrease is much more slowly for high permittivity values $\kappa_b \gg \kappa$ than for the small ones $\kappa_b \ll \kappa$ (compare Figs.9 a-d).

The invariant $F_3(q)$ determines PFM resolution for a sinusoidal structure with period $\pi/q$ allowing for the linearity of imaging theory. Thus we predict that PFM image of domain structure in thin ferroelectric films on substrates with high dielectric permittivity $\kappa_b \geq \kappa$ (e.g. SrTiO$_3$, SrRuO$_3$ or metal) should be more visible than for low-$\kappa_b$ ones, whereas the transfer function *flattening* effect (i.e. extension towards high wave vectors) with thickness decrease are almost independent on $\kappa_b$ value. Moreover, the flattening effect responsible for the image sharpness increase with thickness decrease exists in the case $\kappa_b = \kappa$ (see Fig.9b), proving that it is related mainly with the change of the mechanical conditions. Namely, for homogeneous semi-space the external electric field as well as the induced piezoelectric stress is inhomogeneous, thus the PFM response volume is effectively clamped by the surrounding media (3D self-clamping). For the ultrathin films the external field is practically the same across the thickness of PFM response volume, and the clamping becomes 2D (lateral), similarly to misfit strain in epitaxial films.

The changes in the electrostatic potential structure related with substrate permittivity $\kappa_b$ value can decrease or increase the overall amplitude of the transfer function, $F_3(q)$, but its halfwidth remains almost constant. The resolution $r_{min}$ in PFM experiments is determined by the inverse halfwidth of $F_3(q)$. To illustrate the OTF half-width independency on $\kappa_b$, the dependence of Rayleigh two-point resolution $r_{min}$ on PbTiO$_3$ film thickness $h/d$ is shown in Fig.10 for different $\kappa_b$.

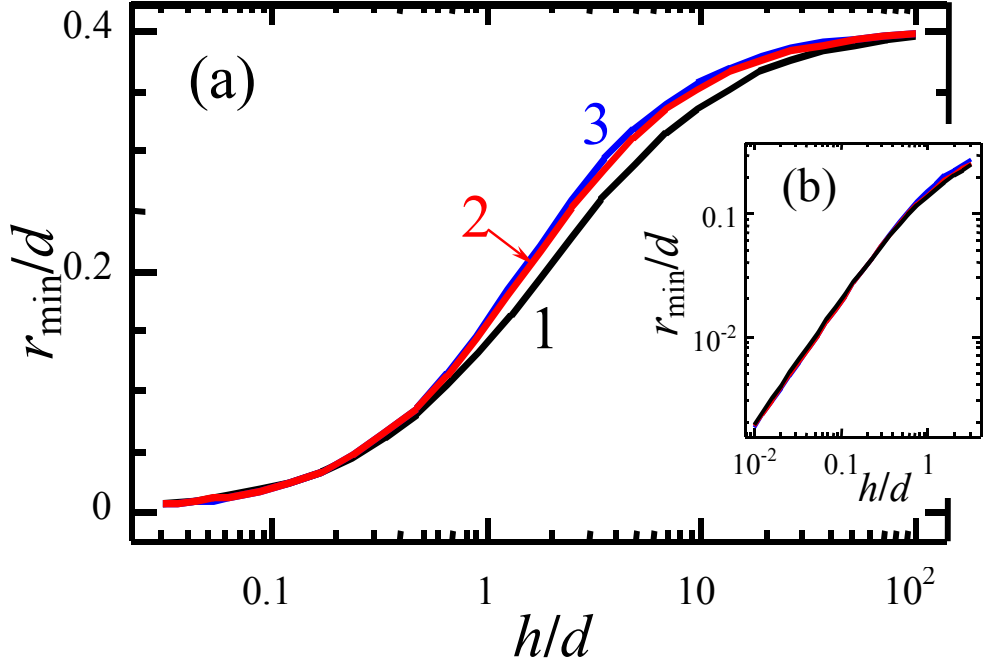

FIG 10. (Color online). The dependence of two-point resolution $r_{min}$ on PbTiO$_3$ film thickness for different values of permittivity $\kappa_b = 3; 300; \infty$ (curves 1, 2, 3).

It is clear that $r_{min}$ is almost independent on $\kappa_b$ value; it decreases with the film thickness decrease, in particular $r_{min} \sim h/2$ at $h \ll \gamma d$ (see Fig.10b). Obviously, the signal strength essentially decreases with film thickness decrease, eventually making the noise level relatively higher. Hypothetically, measuring two-point resolution $r_{min}$ (and knowing that effective distance $d \gg h$) it is possible to determine the thickness $h$ of the thin surface piezoelectric layer. However, ultrathin films provide an effective strategy for achieving high lateral resolutions required for data storage, provided that the sufficiently high sensitivity and low noise level of the detection system can be achieved.

**Discussion**

The effective piezoelectric response, object transfer function components and Rayleigh two-point resolution of piezoelectric surface layers and thin films on rigid and elastically matched substrates are studied. Obtained exact series and simple Pade approximations can be applied for analytical calculations of effective piezoresponse for the films capped on rigid substrates or substrates with matched elastic properties. The following key points are derived.

(a) We predict that effective piezoresponse is thickness dependent for piezoelectric films on substrates with low dielectric permittivity (extrinsic size effect), whereas the thickness dependence is essentially suppressed for giant permittivity or metallic substrates. The fact was explained by the changes of the probe electric field structure inside the film: both the strength and homogeneity of the field increases with substrate permittivity increase.

(b) The transfer function amplitude decreases with thickness decrease. The decrease is suppressed for substrates with high permittivity values and very significant for the small permittivity values.

(c) The transfer function flattening (i.e. extension towards high wave vectors) effect with thickness increase is responsible for the image sharpness increase. It is related mainly with the change of the mechanical clamping conditions with thickness decrease, i.e. cross-over from 3D clamping to 2D clamping. As a result the Rayleigh two-point resolution is almost independent on substrate permittivity and scales linearly with the film thickness.

From the above statements one can conclude that thin piezoelectric films on the metallic ($\kappa_b \to \infty$) or giant permittivity (e.g. relaxor with $\kappa_b \sim 10^4 - 10^5$) substrates are promising objects for PFM visualization of intrinsic size effects (related with the inhomogeneous distribution of the order parameter on the film thickness) since the studied extrinsic ones are suppressed. Several factors may hinder this possibility, e.g. the finiteness of the electric field penetration depth of the metallic/semiconductor substrate could lead to the potential drop; the increase of breakdown probability for the ultrathin films, which may cause the probe damage and subsequent decrease of visibility of domain structure with PFM. Despite these warnings, thin films on giant permittivity substrates could be promising systems for ultrahigh density ferroelectric storage providing sufficiently high-sensitivity and achieving low-noise detection.

**Acknowledgements**

Research supported in part (SVK) by Division of Materials Science and Engineering, Oak Ridge National Laboratory, managed by UT-Battelle, LLC, for the U.S. Department of Energy under Contract DE-AC05-00OR22725. The authors thank Professor Maya D. Glinchuk (Institute for Problems of Materials Science, NAS of Ukraine) for valuable remarks.

**APPENDIX A.**

**I. Fourier representation for Green's function.**

Here we derive the elastic Green's function for the layer on the rigid substrate. General equation for the field of elastic displacement vector is (see e.g. Ref. [35])

$$\Delta_x \mathbf{u} + \frac{1}{1-2\nu} grad_x(div_x \mathbf{u}) = -\frac{2(1+\nu)}{Y} \boldsymbol{F} \cdot \delta(\mathbf{x} - \boldsymbol{\xi}) \qquad \text{(A.1a)}$$

Here vector $\mathbf{x}$ denotes the location of interest and $\boldsymbol{\xi}$ is the point at which the point force, $\boldsymbol{F}$, is applied. The material is isotropic and $\nu$ is Poisson coefficient, $Y$ is Young modulus. Introducing shear modulus $\mu = Y/(2(1+\nu))$, Eq. (A.1a) can be rewritten as:

$$\frac{\partial^2 u_i}{\partial x_k \partial x_k} + \frac{1}{1-2\nu}\frac{\partial^2 u_m}{\partial x_i \partial x_m} = -\frac{F_i}{\mu}\delta(\mathbf{x}-\boldsymbol{\xi}) \qquad (A.1b)$$

Introducing transversal Fourier transformation

$$\tilde{u}_i(k_1,k_2,x_3) = \frac{1}{2\pi}\int_{-\infty}^{\infty} dx_1 \int_{-\infty}^{\infty} dx_2\, \exp(ik_1 x_1 + ik_2 x_2)\cdot u_i(\mathbf{x}), \qquad (A.2a)$$

$$u_i(\mathbf{x}) = \frac{1}{2\pi}\int_{-\infty}^{\infty} dk_1 \int_{-\infty}^{\infty} dk_2\, \exp(-ik_1 x_1 - ik_2 x_2)\, \tilde{u}_i(k_1,k_2,x_3), \qquad (A.2b)$$

and using integral representation of the delta function

$$\delta(x_1-\xi_1)\delta(x_2-\xi_2) = \frac{1}{(2\pi)^2}\int_{-\infty}^{\infty}\int_{-\infty}^{\infty} dk_1 dk_2\, \exp(-ik_1(x_1-\xi_1)-ik_2(x_2-\xi_2)). \qquad (A.3)$$

Eq. (A.1b) yields:

$$\begin{cases} -\left(k_1^2(1+\alpha)+k_2^2\right)\tilde{u}_1 + \dfrac{\partial^2 \tilde{u}_1}{\partial x_3^2} - \alpha k_1 k_2 \tilde{u}_2 - i\alpha k_1 \dfrac{\partial \tilde{u}_3}{\partial x_3} = -\dfrac{F_1}{2\pi\mu}\exp(ik_1\xi_1 + ik_2\xi_2)\delta(x_3-\xi_3) \\[2mm] -\alpha k_1 k_2 \tilde{u}_1 - \left(k_1^2 + k_2^2(1+\alpha)\right)\tilde{u}_2 + \dfrac{\partial^2 \tilde{u}_2}{\partial x_3^2} - i\alpha k_2 \dfrac{\partial \tilde{u}_3}{\partial x_3} = -\dfrac{F_2}{2\pi\mu}\exp(ik_1\xi_1 + ik_2\xi_2)\delta(x_3-\xi_3) \quad (A.4)\\[2mm] -i\alpha k_1 \dfrac{\partial \tilde{u}_1}{\partial x_3} - i\alpha k_2 \dfrac{\partial \tilde{u}_2}{\partial x_3} - \left(k_1^2+k_2^2\right)\tilde{u}_3 + (1+\alpha)\dfrac{\partial^2 \tilde{u}_3}{\partial x_3^2} = -\dfrac{F_3}{2\pi\mu}\exp(ik_1\xi_1 + ik_2\xi_2)\delta(x_3-\xi_3) \end{cases}$$

where $\alpha = 1/(1-2\nu)$ is introduced.

We seek the solution of Eq. (A.4) as $\tilde{u}_i \sim \exp(s x_3)$, where $s$ values should be determined. Substitution into Eq. (A.4) with $F_i = 0$ (homogeneous system) yields the characteristic equation for $s$:

$$-\left(k_1^2 + k_2^2 - s^2\right)^3 (1+\alpha) = 0. \qquad (A.5)$$

Eq. (A5) has triply degenerated roots $s = \pm k$, where $k = \sqrt{k_1^2 + k_2^2}$ is the module of vector $\mathbf{k}$. Hence, the general homogeneous solution of Eq. (A.4) has to be reconstructed from $\exp(\pm k\, x_3)$, $x_3 \exp(\pm k\, x_3)$ and $x_3^2 \exp(\pm k\, x_3)$. After the simple, but cumbersome transformations one can find the general homogeneous solution of Eq. (A.4) as

$$\tilde{u}_1^h(k_1,k_2,x_3) = (C_{10} + i k_1 x_3 C_{31})\exp(-k\, x_3) + (C_{11} + i k_1 x_3 C_{33})\exp(k\, x_3), \quad \text{(A.6a)}$$

$$\tilde{u}_2^h(k_1,k_2,x_3) = (C_{20} + i k_2 x_3 C_{31})\exp(-k\, x_3) + (C_{21} + i k_2 x_3 C_{33})\exp(k\, x_3), \quad \text{(A.6b)}$$

$$\tilde{u}_3^h(k_1,k_2,x_3) = \left(-i\frac{k_1}{k}C_{10} - i\frac{k_2}{k}C_{20} + (3-4\nu)C_{31} + k\, x_3 C_{31}\right)\exp(-k\, x_3) + \\ + \left(i\frac{k_1}{k}C_{11} + i\frac{k_2}{k}C_{21} + (3-4\nu)C_{33} - k\, x_3 C_{33}\right)\exp(k\, x_3) \quad \text{(A.6c)}$$

Note, that we used the equality $1 + 2/\alpha = 3 - 4\nu$ in Eqs.(A.6).

To complement the general solution, we seek the particular solution $u_i^p(\mathbf{x})$ of the inhomogeneous Eqs.(A.4). One of the simplest is the solution for the homogeneous space, $u_i^\infty(\mathbf{x})$. Using the full 3D-Fourier transformation:

$$\hat{u}_i^\infty(\mathbf{k}) = \frac{1}{(2\pi)^{3/2}} \int_{-\infty}^{\infty} dx_1 \int_{-\infty}^{\infty} dx_2 \int_{-\infty}^{\infty} dx_3\, \exp(i\mathbf{kx})\cdot u_i^\infty(\mathbf{x}), \quad \text{(A.7a)}$$

$$u_i^\infty(\mathbf{x}) = \frac{1}{(2\pi)^{3/2}} \int_{-\infty}^{\infty} dk_1 \int_{-\infty}^{\infty} dk_2 \int_{-\infty}^{\infty} dk_3\, \exp(-i\mathbf{kx})\cdot \hat{u}_i^\infty(\mathbf{k}), \quad \text{(A.7b)}$$

Eq. (A.1) is reduced to the system of algebraic equations:

$$\begin{pmatrix} k_1^2(1+\alpha)+k_2^2+k_3^2 & \alpha k_1 k_2 & \alpha k_1 k_3 \\ \alpha k_1 k_2 & k_1^2 + k_2^2(1+\alpha) + k_3^2 & \alpha k_2 k_3 \\ \alpha k_1 k_3 & \alpha k_2 k_3 & k_1^2 + k_2^2 + k_3^2(1+\alpha) \end{pmatrix} \begin{pmatrix} \hat{u}_1^\infty \\ \hat{u}_2^\infty \\ \hat{u}_3^\infty \end{pmatrix} = \frac{\exp(i\mathbf{k}\boldsymbol{\xi})}{(2\pi)^{\frac{3}{2}}\mu} \begin{pmatrix} F_1 \\ F_2 \\ F_3 \end{pmatrix} \quad \text{(A.8)}$$

Its solution has the form:

$$\hat{u}_i^\infty(\mathbf{k}) = \hat{G}_{ij}^\infty(\mathbf{k})F_j \exp(i\mathbf{k}\boldsymbol{\xi}), \quad \hat{G}_{ij}^\infty(\mathbf{k}) = \frac{1}{(2\pi)^{3/2}\mu}\left(\frac{\delta_{ij}}{|\mathbf{k}|^2} - \frac{1}{2(1-\nu)}\frac{k_i k_j}{|\mathbf{k}|^4}\right) \quad \text{(A.9)}$$

The inhomogeneous solution (A.9) corresponds to the well-known Fourier image of Green's tensor for infinite homogeneous isotropic media (see e.g. Ref. 36). Since we are looking for solution of system, confined in one direction ($x_3$), it is convenient to transform (A.9) to coordinate representation on $x_3$. Simple integration gives

$$\tilde{u}_i^\infty(k_1,k_2,x_3) = \tilde{G}_{ij}^\infty(k_1,k_2,x_3-\xi_3)F_j \exp(ik_1\xi_1 + ik_2\xi_2), \quad (A.10)$$

where

$$\tilde{G}_{11}^\infty(k_1,k_2,x_3-\xi_3) = \frac{1}{4\pi\mu}\frac{\exp(-k|x_3-\xi_3|)}{k}\left(1 - \frac{k_1^2(1+k|x_3-\xi_3|)}{4(1-\nu)k^2}\right) \quad (A.11a)$$

$$\tilde{G}_{12}^\infty(k_1,k_2,x_3-\xi_3) = -\frac{1}{4\pi\mu}\frac{\exp(-k|x_3-\xi_3|)}{k}\frac{k_1 k_2(1+k|x_3-\xi_3|)}{4(1-\nu)k^2} \quad (A.11b)$$

$$\tilde{G}_{22}^\infty(k_1,k_2,x_3-\xi_3) = \frac{1}{4\pi\mu}\frac{\exp(-k|x_3-\xi_3|)}{k}\left(1 - \frac{k_2^2(1+k|x_3-\xi_3|)}{4(1-\nu)k^2}\right) \quad (A.11c)$$

$$\tilde{G}_{13}^\infty(k_1,k_2,x_3-\xi_3) = \frac{1}{4\pi\mu}\frac{\exp(-k|x_3-\xi_3|)}{k}\frac{ik_1(x_3-\xi_3)}{4(1-\nu)} \quad (A.11d)$$

$$\tilde{G}_{23}^\infty(k_1,k_2,x_3-\xi_3) = \frac{1}{4\pi\mu}\frac{\exp(-k|x_3-\xi_3|)}{k}\frac{ik_2(x_3-\xi_3)}{4(1-\nu)} \quad (A.11e)$$

$$\tilde{G}_{33}^\infty(k_1,k_2,x_3-\xi_3) = \frac{1}{4\pi\mu}\frac{\exp(-k|x_3-\xi_3|)}{k}\left(1 - \frac{1-k|x_3-\xi_3|}{4(1-\nu)}\right) \quad (A.11f)$$

and $\mu = Y/(2(1+\nu))$.

The general solution of the chosen elastic problem should satisfy the boundary conditions at the rigid substrate ($u_i(x_3=h)=0$) and free upper surface ($\sigma_{3i}(x_3=0)=0$). Keeping in mind that $\sigma_{ij} = c_{ijkl}u_{kl}$, where $u_{kl} = \frac{1}{2}\left(\frac{\partial u_k}{\partial x_l} + \frac{\partial u_l}{\partial x_k}\right)$ and

$$c_{ijkl} = \frac{Y}{2(1+\nu)}\left(\frac{2\nu}{1-2\nu}\delta_{ij}\delta_{kl} + \delta_{ik}\delta_{jl} + \delta_{il}\delta_{jk}\right),$$ one obtains that:

$$\sigma_{33} = \frac{Y}{(1+\nu)(1-2\nu)}((1-\nu)u_{33} + \nu(u_{11}+u_{22})), \quad \sigma_{31} = \frac{Y}{(1+\nu)}u_{13}, \quad \sigma_{32} = \frac{Y}{(1+\nu)}u_{23}.$$

Finally, the boundary conditions for $\tilde{u}_i(k_1,k_2,x_3) = \tilde{u}_i^h(k_1,k_2,x_3) + \tilde{u}_i^p(k_1,k_2,x_3)$ have the form:

$$\begin{cases} (1-\nu)\dfrac{\partial \tilde{u}_3}{\partial x_3} - i\nu(k_1\tilde{u}_1 + k_2\tilde{u}_2)\Big|_{x_3=0} = 0, \quad \dfrac{\partial \tilde{u}_1}{\partial x_3} - ik_1\tilde{u}_3\Big|_{x_3=0} = 0, \quad \dfrac{\partial \tilde{u}_2}{\partial x_3} - ik_2\tilde{u}_3\Big|_{x_3=0} = 0, \\ \tilde{u}_1\big|_{x_3=h} = 0, \quad \tilde{u}_2\big|_{x_3=h} = 0, \quad \tilde{u}_3\big|_{x_3=h} = 0. \end{cases} \quad (A.12)$$

Six constants $C_{ij}$ should be expressed via $F_j$ from Eqs.(A.12) as solution for

$$(C_{10} + ik_1 h C_{31})\exp(-kh) + (C_{11} + ik_1 h C_{33})\exp(kh) = -\tilde{u}_1^p(k_1,k_2,h), \quad (A.13a)$$

$$(C_{20} + ik_2 h C_{31})\exp(-kh) + (C_{21} + ik_2 h C_{33})\exp(kh) = -\tilde{u}_2^p(k_1,k_2,h) \quad (A.13b)$$

$$\left(\begin{array}{l}\left(-i\dfrac{k_2}{k}C_{20} - i\dfrac{k_1}{k}C_{10} + (3-4\nu)C_{31} + khC_{31}\right)\exp(-kh) + \\ +\left(i\dfrac{k_2}{k}C_{21} + i\dfrac{k_1}{k}C_{11} + (3-4\nu)C_{33} - khC_{33}\right)\exp(kh)\end{array}\right) = -\tilde{u}_3^p(k_1,k_2,h) \quad (A.13c)$$

$$\begin{pmatrix}-kC_{10}+ik_1C_{31}+\\+kC_{11}+ik_1C_{33}\end{pmatrix} - ik_1\begin{pmatrix}-i\dfrac{k_2}{k}C_{20}-i\dfrac{k_1}{k}C_{10}+(3-4\nu)C_{31}+\\+i\dfrac{k_2}{k}C_{21}+i\dfrac{k_1}{k}C_{11}+(3-4\nu)C_{33}\end{pmatrix} = \begin{pmatrix}-\dfrac{\partial \tilde{u}_1^p(k_1,k_2,0)}{\partial x_3}+\\+ik_1\tilde{u}_3^p(k_1,k_2,0)\end{pmatrix} \quad (A.13d)$$

$$\begin{pmatrix}-kC_{20}+ik_2C_{31}+\\+kC_{21}+ik_2C_{33}\end{pmatrix} - ik_2\begin{pmatrix}-i\dfrac{k_2}{k}C_{20}-i\dfrac{k_1}{k}C_{10}+(3-4\nu)C_{31}+\\+i\dfrac{k_2}{k}C_{21}+i\dfrac{k_1}{k}C_{11}+(3-4\nu)C_{33}\end{pmatrix} = \begin{pmatrix}-\dfrac{\partial \tilde{u}_2^p(k_1,k_2,0)}{\partial x_3}+\\+ik_2\tilde{u}_3^p(k_1,k_2,0)\end{pmatrix} \quad (A.13e)$$

$$(1-\nu)k\begin{pmatrix}i\dfrac{k_2}{k}C_{21}+i\dfrac{k_1}{k}C_{11}+(3-4\nu)C_{33}-C_{33}+\\ +i\dfrac{k_2}{k}C_{20}+i\dfrac{k_1}{k}C_{10}-(3-4\nu)C_{31}+C_{31}\end{pmatrix}-i\nu\begin{pmatrix}k_1(C_{10}+C_{11})+\\ +k_2(C_{20}+C_{21})\end{pmatrix}=\begin{pmatrix}-(1-\nu)\dfrac{\partial\tilde{u}_3^p(k_1,k_2,0)}{\partial x_3}\\ +i\nu\begin{pmatrix}k_1\tilde{u}_1^p(k_1,k_2,0)+\\ +k_2\tilde{u}_3^p(k_1,k_2,0)\end{pmatrix}\end{pmatrix}$$

(A.13f)

(I) <u>First step</u>. Let us find the Green function $\tilde{G}_{ij}^s(k_1,k_2,x_3-\xi_3)$ of the semi-space ($h\to\infty$). For the case $C_{11}=C_{21}=C_{33}=0$ and:

$$\tilde{u}_1^h(k_1,k_2,x_3)=(C_{10}+ik_1x_3C_{31})\exp(-kx_3),\qquad (A.14a)$$

$$\tilde{u}_2^h(k_1,k_2,x_3)=(C_{20}+ik_2x_3C_{31})\exp(-kx_3),\qquad (A.14b)$$

$$\tilde{u}_3^h(k_1,k_2,x_3)=\left(-i\dfrac{k_1}{k}C_{10}-i\dfrac{k_2}{k}C_{20}+(3-4\nu)C_{31}+kx_3C_{31}\right)\exp(-kx_3).\quad (A.14c)$$

Then Eqs. (A.13d-f) should be solved under the condition $\tilde{u}_i^p(k_1,k_2,0)\equiv\tilde{u}_i^\infty(k_1,k_2,0)$. After cumbersome algebraic transformations we derive:

$$\tilde{u}_i^s(k_1,k_2,x_3)=\tilde{G}_{ij}^s(k_1,k_2,x_3,\xi_3)F_j\exp(ik_1\xi_1+ik_2\xi_2).\qquad (A.15)$$

Where

$$\tilde{G}_{11}^s(k_1,k_2,x_3,\xi_3)=\dfrac{\exp(-k(x_3+\xi_3))}{16\pi\mu k^3(1-\nu)}\begin{pmatrix}-k_1^2k(x_3+\xi_3)(3-4\nu)+\\ +2k^2(2-2\nu+k_1^2x_3\xi_3)+k_1^2(1-8(1-\nu)\nu)\end{pmatrix}+\\ +\dfrac{\exp(-k|x_3-\xi_3|)}{16\pi\mu k^3(1-\nu)}(4k^2(1-\nu)-k_1^2-k_1^2k|x_3-\xi_3|)$$

(A.16a)

$$\tilde{G}_{21}^s(k_1,k_2,x_3,\xi_3)=\dfrac{\exp(-k(x_3+\xi_3))}{16\pi\mu k^3(1-\nu)}k_1k_2\begin{pmatrix}2k^2x_3\xi_3-k(x_3+\xi_3)(3-4\nu)+\\ +1-8(1-\nu)\nu\end{pmatrix}+\\ +\dfrac{\exp(-k|x_3-\xi_3|)}{16\pi\mu k^3(1-\nu)}k_1k_2(1+k|x_3-\xi_3|)$$

(A.16b)

$$\widetilde{G}^s_{31}(k_1,k_2,x_3,\xi_3) = \frac{\exp(-k(x_3+\xi_3))}{16\pi\mu k^2(1-\nu)} ik_1 \begin{pmatrix} -2k^2 x_3\xi_3 + k(x_3-\xi_3)(3-4\nu)+ \\ +4(1-\nu)(1-2\nu) \end{pmatrix} +$$
$$+ \frac{\exp(-k|x_3-\xi_3|)}{16\pi\mu k(1-\nu)} ik_1 (x_3-\xi_3) \quad (A.16c)$$

$$\widetilde{G}^s_{12}(k_1,k_2,x_3,\xi_3) = \frac{\exp(-k(x_3+\xi_3))}{16\pi\mu k^3(1-\nu)} k_1 k_2 \begin{pmatrix} 2k^2 x_3\xi_3 - k(x_3+\xi_3)(3-4\nu)+ \\ +1-8(1-\nu)\nu \end{pmatrix} +$$
$$+ \frac{\exp(-k|x_3-\xi_3|)}{16\pi\mu k^3(1-\nu)} k_1 k_2 (1+k|x_3-\xi_3|) \quad (A.16d)$$

$$\widetilde{G}^s_{22}(k_1,k_2,x_3,\xi_3) = \frac{\exp(-k(x_3+\xi_3))}{16\pi\mu k^3(1-\nu)} \begin{pmatrix} -k_2^2 k(x_3+\xi_3)(3-4\nu)+ \\ +2k^2(2-2\nu+k_2^2 x_3\xi_3) + k_2^2(1-8(1-\nu)\nu) \end{pmatrix} +$$
$$+ \frac{\exp(-k|x_3-\xi_3|)}{16\pi\mu k^3(1-\nu)} (4k^2(1-\nu) - k_2^2 - k_2^2 k|x_3-\xi_3|) \quad (A.16e)$$

$$\widetilde{G}^s_{32}(k_1,k_2,x_3,\xi_3) = \frac{\exp(-k(x_3+\xi_3))}{16\pi\mu k^2(1-\nu)} ik_2 \begin{pmatrix} -2k^2 x_3\xi_3 + k(x_3-\xi_3)(3-4\nu)+ \\ +4(1-\nu)(1-2\nu) \end{pmatrix} +$$
$$+ \frac{\exp(-k|x_3-\xi_3|)}{16\pi\mu k(1-\nu)} ik_2 (x_3-\xi_3) \quad (A.16f)$$

$$\widetilde{G}^s_{13}(k_1,k_2,x_3,\xi_3) = \frac{\exp(-k(x_3+\xi_3))}{16\pi\mu k^2(1-\nu)} ik_1 \begin{pmatrix} 2k^2 x_3\xi_3 + k(x_3-\xi_3)(3-4\nu)+ \\ -4(1-\nu)(1-2\nu) \end{pmatrix} +$$
$$+ \frac{\exp(-k|x_3-\xi_3|)}{16\pi\mu k(1-\nu)} ik_1 (x_3-\xi_3) \quad (A.16g)$$

$$\widetilde{G}^s_{23}(k_1,k_2,x_3,\xi_3) = \frac{\exp(-k(x_3+\xi_3))}{16\pi\mu k^2(1-\nu)} ik_2 \begin{pmatrix} 2k^2 x_3\xi_3 + k(x_3-\xi_3)(3-4\nu)- \\ -4(1-\nu)(1-2\nu) \end{pmatrix} +$$
$$+ \frac{\exp(-k|x_3-\xi_3|)}{16\pi\mu k(1-\nu)} ik_2 (x_3-\xi_3) \quad (A.16h)$$

$$\widetilde{G}^s_{33}(k_1,k_2,x_3,\xi_3) = \frac{\exp(-k(x_3+\xi_3))}{16\pi\mu k(1-\nu)} \begin{pmatrix} 2k^2 x_3\xi_3 + k(x_3+\xi_3)(3-4\nu)+ \\ +1+4(1-\nu)(1-2\nu) \end{pmatrix} +$$
$$+ \frac{\exp(-k|x_3-\xi_3|)}{16\pi\mu k(1-\nu)} (3-4\nu + k|x_3-\xi_3|) \quad (A.16i)$$

where $k = \sqrt{k_1^2 + k_2^2}$. Note that $\tilde{G}_{21}^s(k_1,k_2,x_3,\xi_3) = \tilde{G}_{12}^s(k_1,k_2,x_3,\xi_3)$, $\tilde{G}_{22}^s(k_1,k_2,x_3,\xi_3) = \tilde{G}_{11}^s(k_2,k_1,x_3,\xi_3)$, $\tilde{G}_{32}^s(k_1,k_2,x_3,\xi_3) = \tilde{G}_{31}^s(k_2,k_1,x_3,\xi_3)$ as expected.

(II) <u>Second step</u>. Using Eq.(A.15) as the partial solution $\tilde{u}_i^p(k_1,k_2,x_3) = \tilde{u}_i^s(k_1,k_2,x_3)$, let us find the surface vertical displacement $\tilde{u}_3^f(k_1,k_2,0) = \tilde{u}_3^h(k_1,k_2,0) + \tilde{u}_3^s(k_1,k_2,0)$ for the film of thickness $h$. Here $\tilde{u}_3^h(k_1,k_2,0)$ should be found from Eqs.(A.13), namely:

$$i\frac{k_1}{k}(C_{10} - C_{11}) + i\frac{k_2}{k}(C_{20} - C_{21}) - (3 - 4\nu)(C_{31} + C_{33}) + \tilde{u}_3^h(k_1,k_2,0) = 0, \quad (A.17a)$$

$$C_{10} - C_{11} - \frac{ik_1}{k}(C_{31} + C_{33}) + \frac{ik_1}{k}\tilde{u}_3^h(k_1,k_2,0) = 0, \quad (A.17b)$$

$$C_{20} - C_{21} - \frac{ik_2}{k}(C_{31} + C_{33}) + \frac{ik_2}{k}\tilde{u}_3^h(k_1,k_2,0) = 0, \quad (A.17c)$$

$$i\frac{k_1}{k}(1 - 2\nu)(C_{10} + C_{11}) + i\frac{k_2}{k}(1 - 2\nu)(C_{20} + C_{21}) - (1 - \nu)(2 - 4\nu)(C_{31} - C_{33}) = 0, \quad (A.17d)$$

$$C_{10}\exp(-kh) + C_{11}\exp(kh) + ik_1 h(C_{31}\exp(-kh) + C_{33}\exp(kh)) = -\tilde{u}_1^s(k_1,k_2,h), \quad (A.17e)$$

$$C_{20}\exp(-kh) + C_{21}\exp(kh) + ik_2 h(C_{31}\exp(-kh) + C_{33}\exp(kh)) = -\tilde{u}_2^s(k_1,k_2,h), \quad (A.17f)$$

$$\left( \begin{array}{l} -i\dfrac{k_1}{k}C_{10}\exp(-kh) + i\dfrac{k_1}{k}C_{11}\exp(kh) - \\ -i\dfrac{k_2}{k}C_{20}\exp(-kh) + i\dfrac{k_2}{k}C_{21}\exp(kh) + \\ + (3 - 4\nu + kh)C_{31}\exp(-kh) + (3 - 4\nu - kh)C_{33}\exp(kh) \end{array} \right) = -\tilde{u}_3^s(k_1,k_2,h). \quad (A.17g)$$

After cumbersome algebraic transformations we derived that

$$\tilde{u}_3^f(k_1,k_2,0) = \tilde{u}_3^s(k_1,k_2,0) - \frac{4(1-\nu)\tilde{u}_3^s(k_1,k_2,h)(\exp(-kh)(2 - 2\nu - kh) + \exp(kh)(2 - 2\nu + kh))}{(3 - 4\nu)(\exp(-2kh) + \exp(2kh)) + 2(1 + 4(1-\nu)(1-2\nu) + 2k^2h^2)} +$$

$$+ \frac{4(1-\nu)i(k_1\tilde{u}_1^s(k_1,k_2,h) + k_2\tilde{u}_2^s(k_1,k_2,h))(\exp(-kh)(1 - 2\nu + kh) - \exp(kh)(1 - 2\nu - kh))}{k((3 - 4\nu)(\exp(-2kh) + \exp(2kh)) + 2(1 + 4(1-\nu)(1-2\nu) + 2k^2h^2))}$$

(A.18)

Finally, in agreement with Eq.(A.15) one obtains

$$\tilde{u}_3^f(k_1,k_2,0) = \tilde{G}_{3j}^f(k_1,k_2,\xi_3)F_j \exp(ik_1\xi_1 + ik_2\xi_2). \qquad (A.19)$$

Where the elastic Green function $\tilde{G}_{3j}^f(k_1,k_2,\xi_3)$ for the film on a rigid substrate has the form:

$$\tilde{G}_{3j}^f(k_1,k_2,\xi_3) = \begin{pmatrix} \tilde{G}_{3j}^s(k_1,k_2,0,\xi_3) - \tilde{G}_{3j}^s(k_1,k_2,h,\xi_3)\phi_3(k\,h,\nu) + \\ + i(k_1\tilde{G}_{1j}^s(k_1,k_2,h,\xi_3) + k_2\tilde{G}_{2j}^s(k_1,k_2,h,\xi_3))\phi_\perp(k\,h,\nu) \end{pmatrix}, \qquad (A.20)$$

$$\phi_3(k\,h,\nu) = \frac{4(1-\nu)(\exp(-k\,h)(2-2\nu-k\,h) + \exp(k\,h)(2-2\nu+k\,h))}{(3-4\nu)(\exp(-2k\,h) + \exp(2k\,h)) + 2(1+4(1-\nu)(1-2\nu)+2k^2h^2)}, \qquad (A.20b)$$

$$\phi_\perp(k\,h,\nu) = \frac{4(1-\nu)(\exp(-k\,h)(1-2\nu+k\,h) - \exp(k\,h)(1-2\nu-k\,h))}{k((3-4\nu)(\exp(-2k\,h) + \exp(2k\,h)) + 2(1+4(1-\nu)(1-2\nu)+2k^2h^2))}, \qquad (A.20c)$$

Here $k \equiv \sqrt{k_1^2 + k_2^2}$, $\nu$ is the Poisson ratio. Note that $\tilde{G}_{3j}^f(k_1,k_2,0<\xi_3<h) = 0$ at $h=0$ as it should be expected. Simplified, but qualitatively correct expression is

$$\tilde{G}_{3j}^f(k_1,k_2,\xi_3) \approx \tilde{G}_{3j}^s(k_1,k_2,0,\xi_3) - \tilde{G}_{3j}^s(k_1,k_2,h,\xi_3)\cosh(k\,h)\exp(-2k\,h) - $$
$$-\frac{(1-2\nu)}{2k(1-\nu)}i(k_1\tilde{G}_{1j}^s(k_1,k_2,h,\xi_3) + k_2\tilde{G}_{2j}^s(k_1,k_2,h,\xi_3))\sinh(k\,h)\exp(-2k\,h) \qquad (A.21)$$

**II. Object transfer function (OTF).**

In the case when $x_1 = x_2 = 0$ (the displacement directly below the tip) and strain piezoelectric coefficient $d_{klj}(\xi)$ is independent on $\xi_3$ (system is uniform within the layer in z-direction), the vertical surface displacement $u_3(0,\mathbf{y})$ is given by the convolution of stress piezoelectric tensor components $d_{klj}(\mathbf{y}-\xi)$ representing material properties (*ideal image*) with the resolution function components $W_{3jkl}(\xi)$. The Fourier transformation of vertical surface displacement $\tilde{u}_3(q_1,q_2)$ is:

$$\tilde{u}_3(q_1,q_2) = \tilde{d}_{klj}(q_1,q_2)\tilde{W}_{3jkl}(-q_1,-q_2), \qquad (A.22)$$

where the resolution function Fourier image (Object Transfer Function) that contributes into vertical displacement $u_3$ is introduced as

$$\tilde{W}_{3jkl}(q_1,q_2) = \int_{-\infty}^{\infty}dk_1\int_{-\infty}^{\infty}dk_2\int_0^h d\xi_3 \tilde{G}^f_{3m,n}(q_1-k_1,q_2-k_2,\xi_3)c_{kjmn}\tilde{E}_l(k_1,k_2,\xi_3) \qquad (A.23a)$$

$$\tilde{G}^f_{3m,n}(k_1,k_2,\xi_3) \equiv \begin{cases} ik_n\tilde{G}^f_{3m}(k_1,k_2,\xi_3), & n=1,2 \\ \dfrac{\partial}{\partial\xi_3}\tilde{G}^f_{3m}(k_1,k_2,\xi_3), & n=3 \end{cases} \qquad (A.23b)$$

The electric field $E_l(x_1,x_2,\xi_3)$ is produced by the tip in the point $\mathbf{x}=(x_1,x_2,z)$ of the sample, $c_{kjmn}$ are stiffness tensor components.

The number of non-zero components of OTF depends on material symmetry and for transversally isotropic materials (e.g. tetragonal perovskites) only the components $\tilde{W}_{333,313,351}$ are non-zero (in Voigt representation). The vertical response in this case is

$$\tilde{u}_3(q) = \tilde{W}_{333}(q)d_{33} + \tilde{W}_{313}(q)d_{31} + \tilde{W}_{351}(q)d_{15}. \qquad (A.24)$$

After elementary, but very cumbersome transformations we obtain:

$$\tilde{W}_{333}(q_1,q_2) = \frac{Q}{2\pi\varepsilon_0}\int_{-\infty}^{\infty}dk_1\int_{-\infty}^{\infty}dk_2\int_0^h d\xi_3 \tilde{G}^f_{3m,n}(q_1-k_1,q_2-k_2,\xi_3)c_{33mn}\tilde{E}_3(k,\xi_3) \qquad (A.25)$$

$$\tilde{W}_{313}(q_1,q_2) = \frac{Q}{2\pi\varepsilon_0}\int_{-\infty}^{\infty}dk_1\int_{-\infty}^{\infty}dk_2\int_0^h d\xi_3 \tilde{G}^f_{3m,n}(q_1-k_1,q_2-k_2,\xi_3)(c_{11mn}+c_{22mn})\tilde{E}_3(k,\xi_3) \qquad (A.26)$$

$$\tilde{W}_{351}(q_1,q_2) = \frac{Q}{2\pi\varepsilon_0}\int_{-\infty}^{\infty}dk_1\int_{-\infty}^{\infty}dk_2\int_0^h d\xi_3 c_{1212}\left(\left(\tilde{G}^f_{33,1}+\tilde{G}^f_{31,3}\right)ik_1 + \left(\tilde{G}^f_{33,2}+\tilde{G}^f_{32,3}\right)ik_2\right)\tilde{\varphi}(k,\xi_3) \qquad (A.27)$$

where $\tilde{E}_3 = -\partial\tilde{\varphi}/\partial\xi_3$, $\tilde{E}_{1,2} = ik_{1,2}\tilde{\varphi}$. The transversely homogeneous film response is

$$\tilde{W}_{333}(0) = \frac{Q}{\varepsilon_0}\int_0^\infty kdk \int_0^h d\xi_3 \tilde{G}^f_{3m,n}(k,\xi_3)c_{33mn}\tilde{E}_3(k,\xi_3) \tag{A.28}$$

$$\tilde{W}_{313}(0) = \frac{Q}{\varepsilon_0}\int_0^\infty kdk \int_0^h d\xi_3 \tilde{G}^f_{3m,n}(k,\xi_3)(c_{11mn}+c_{22mn})\tilde{E}_3(k,\xi_3) \tag{A.29}$$

$$\tilde{W}_{351}(0) = \frac{Q}{\varepsilon_0}\int_0^\infty kdk \int_0^h d\xi_3 c_{1212}\left(\left(\tilde{G}^f_{33,1}+\tilde{G}^f_{31,3}\right)ik_1 + \left(\tilde{G}^f_{33,2}+\tilde{G}^f_{32,3}\right)ik_2\right)\tilde{\varphi}(k,\xi_3) \tag{A.30}$$

$$\begin{aligned}
\tilde{G}^f_{3m,n}(k_1,k_2,\xi_3)c_{33mn} =\ &-\frac{1}{2\pi}\exp(-k\xi_3)(1+k\xi_3) + \\
&+\frac{\exp(-k\xi_3)\left(5-12\nu+8\nu^2+2kh(1+k\xi_3)+2k^2h^2(1+2k\xi_3)+k\xi_3(3-4\nu)^2\right)}{2\pi\left((3-4\nu)(\exp(-2kh)+\exp(2kh))+2(1+4(1-\nu)(1-2\nu)+2k^2h^2)\right)} + \\
&-\frac{\exp(k\xi_3)\left(5-12\nu+8\nu^2+2k^2h^2-k\xi_3+2kh(1-k\xi_3)\right)}{2\pi\left((3-4\nu)(\exp(-2kh)+\exp(2kh))+2(1+4(1-\nu)(1-2\nu)+2k^2h^2)\right)} + \\
&+\frac{(3-4\nu)(\exp(-k(2h+\xi_3))(1+k\xi_3)-\exp(-k(2h-\xi_3))(1-k\xi_3))}{2\pi\left((3-4\nu)(\exp(-2kh)+\exp(2kh))+2(1+4(1-\nu)(1-2\nu)+2k^2h^2)\right)}
\end{aligned} \tag{A.31a}$$

$$\begin{aligned}
\tilde{G}^f_{3m,n}(k_1,k_2,\xi_3)(c_{11mn}+c_{22mn}) =\ &-\frac{1}{2\pi}\exp(-k\xi_3)(1+2\nu-k\xi_3) + \\
&-\frac{\exp(-k\xi_3)\begin{pmatrix}13+18\nu+24\nu^2-32\nu^3-2kh(1+2\nu-k\xi_3)-\\-2k^2h^2(3+4\nu-2k\xi_3)+k\xi_3(3-4\nu)^2\end{pmatrix}}{2\pi\left((3-4\nu)(\exp(-2kh)+\exp(2kh))+2(1+4(1-\nu)(1-2\nu)+2k^2h^2)\right)} + \\
&-\frac{\exp(k\xi_3)\left(-3+14\nu-8\nu^2-2k^2h^2+k\xi_3+2kh(1+2\nu+k\xi_3)\right)}{2\pi\left((3-4\nu)(\exp(-2kh)+\exp(2kh))+2(1+4(1-\nu)(1-2\nu)+2k^2h^2)\right)} + \\
&+\frac{(3-4\nu)(\exp(-k\xi_3)(1+2\nu-k\xi_3)-\exp(k\xi_3)(1+2\nu+k\xi_3))\exp(-2kh)}{2\pi\left((3-4\nu)(\exp(-2kh)+\exp(2kh))+2(1+4(1-\nu)(1-2\nu)+2k^2h^2)\right)}
\end{aligned} \tag{A.31b}$$

$$c_{1212}\left(\left(\widetilde{G}^f_{33,1} + \widetilde{G}^f_{31,3}\right)ik_1 + \left(\widetilde{G}^f_{33,2} + \widetilde{G}^f_{32,3}\right)ik_2\right) = -\frac{\exp(-k\xi_3)k^2\xi_3}{2\pi} +$$

$$-\frac{k\exp(-k\xi_3)\left(4 - 12\nu + 8\nu^2 - 2k^2 h\xi_3 + 2k^2 h^2(1 - 2k\xi_3) - k\xi_3(3 - 4\nu)^2\right)}{2\pi\left((3 - 4\nu)(\exp(-2kh) + \exp(2kh)) + 2(1 + 4(1 - \nu)(1 - 2\nu) + 2k^2 h^2)\right)} +$$

$$+\frac{k\exp(k\xi_3)\left(4 - 12\nu + 8\nu^2 - 2k^2 h\xi_3 + 2k^2 h^2 - k\xi_3\right)}{2\pi\left((3 - 4\nu)(\exp(-2kh) + \exp(2kh)) + 2(1 + 4(1 - \nu)(1 - 2\nu) + 2k^2 h^2)\right)} +$$

$$+\frac{k^2\xi_3(3 - 4\nu)(\exp(-k\xi_3) - \exp(k\xi_3))\exp(-2kh)}{2\pi\left((3 - 4\nu)(\exp(-2kh) + \exp(2kh)) + 2(1 + 4(1 - \nu)(1 - 2\nu) + 2k^2 h^2)\right)}$$

(A.31c)

It is clear that exact expressions (A.31) along with the electric field (7) could be integrated via the coordinate in elementary functions. Thus even for the case of a film clamped on a rigid substrate the response can be obtained in the form of one-fold integrals.

In order to obtain approximate simple analytical expressions the following expansions of (A.31) is derived:

$$\widetilde{G}^f_{3m,n}(k_1,k_2,\xi_3)c_{33mn} = \begin{cases} -\dfrac{(1 - k^2\xi_3^2)}{2\pi} - \dfrac{(1 - 2\nu)k^2\xi_3^2}{4\pi(1 - \nu)} + k^4 O(\xi_3^2 h^2), & kh \ll 1 \\ -\dfrac{\exp(-k\xi_3)(1 + k\xi_3)}{2\pi} - \dfrac{\exp(k\xi_3 - 2kh)}{\pi(3 - 4\nu)}k^2 h(h - \xi_3) + \\ +\dfrac{\exp(-k\xi_3 - 2kh)}{\pi(3 - 4\nu)}k^2 h^2, & kh \gg 1 \end{cases} \quad \text{(A.32a)}$$

$$\widetilde{G}^f_{3m,n}(k_1,k_2,\xi_3)(c_{11mn} + c_{22mn}) = \begin{cases} \dfrac{(1 + \nu)}{\pi}\left(\dfrac{1 - 2\nu}{2(1 - \nu)} - k\xi_3\right) - \dfrac{1 + 2\nu - 2k\xi_3(1 + \nu)}{2\pi}, & kh \ll 1 \\ \dfrac{\exp(-k\xi_3 - 2kh)k^2 h^2(3 + 4\nu)}{\pi(3 - 4\nu)} + \dfrac{\exp(k\xi_3 - 2kh)k^2 h(h - \xi_3)}{\pi(3 - 4\nu)} - \\ -\dfrac{1}{2\pi}\exp(-k\xi_3)(1 + 2\nu - k\xi_3), & kh \gg 1 \end{cases}$$

(A.32b)

$$c_{1212}\left(\left(\tilde{G}^f_{33,1}+\tilde{G}^f_{31,3}\right)ik_1+\left(\tilde{G}^f_{33,2}+\tilde{G}^f_{32,3}\right)ik_2\right)\approx\begin{cases}\dfrac{(1-2\nu)}{4\pi(1-\nu)}k^2\xi_3-\dfrac{k^2\xi_3}{2\pi},& kh\ll 1\\[2mm]\dfrac{k^3 h(h-\xi_3)}{\pi(3-4\nu)}\exp(-2kh+k\xi_3)-\\[2mm]-\dfrac{\exp(-k\xi_3)k^2\xi_3}{2\pi},& kh\gg 1\end{cases}\quad(\text{A.32c})$$

### APPENDIX B. Electric field calculations for film

The external electric field potential $\varphi_i(\mathbf{r})$ created by the point charge $Q$ localized in air in the point $r_0=(0,0,-d)$, inside the film $0\le z\le h$ filled by transversely isotropic dielectric $\varepsilon_i$ could be found from the boundary problem:

$$\begin{aligned}&\Delta\varphi_e(\mathbf{r})=-\frac{Q}{\varepsilon_0\varepsilon_e}\delta(x,y,z+d),\quad z\le 0,\\ &\varepsilon_{11}\Delta_{x,y}\varphi_i(\mathbf{r})+\varepsilon_{33}\frac{\partial^2\varphi_i}{\partial z^2}=0,\quad 0\le z\le h,\\ &\varepsilon^b_{11}\Delta\varphi_b(\mathbf{r})+\varepsilon^b_{33}\frac{\partial^2\varphi_b}{\partial z^2}=0,\quad z\ge h,\\ &\varphi_e(z=0)=\varphi_i(z=0),\quad \left(\varepsilon_e\frac{\partial\varphi_e}{\partial z}-\varepsilon_{33}\frac{\partial\varphi_i}{\partial z}\right)\bigg|_{z=0}=0,\\ &\varphi_i(z=h)=\varphi_b(z=h),\quad \left(\varepsilon^b_{33}\frac{\partial\varphi_b}{\partial z}-\varepsilon_{33}\frac{\partial\varphi_i}{\partial z}\right)\bigg|_{z=h}=0,\end{aligned}\quad(\text{B.1})$$

Note, that $\varphi_i(z=h)=0$ for a conductive substrate. The solution of Eq. (B.1) can be found using Hankel integral transformation:

$$\varphi_e(\mathbf{r})=\frac{Q}{4\pi\varepsilon_0\varepsilon_e}\int_0^\infty dk J_0\left(k\sqrt{x^2+y^2}\right)\left(\exp(-k\cdot|z+d|)+A\exp(k\cdot z)\right)\quad(\text{B.2})$$

$$\varphi_i(\mathbf{r})=\frac{Q}{4\pi\varepsilon_0}\int_0^\infty dk J_0\left(k\sqrt{x^2+y^2}\right)\left(B\exp\left(-k\frac{z}{\gamma}\right)+C\exp\left(k\frac{z}{\gamma}\right)\right)\quad(\text{B.3})$$

$$\varphi_b(\mathbf{r})=\frac{Q}{4\pi\varepsilon_0}\int_0^\infty dk J_0\left(k\sqrt{x^2+y^2}\right)D\exp\left(-k\frac{z}{\gamma_b}\right)\quad(\text{B.4})$$

Here $J_0$ is Bessel function of zero order, $\gamma = \sqrt{\varepsilon_{33}/\varepsilon_{11}}$, $\gamma_b = \sqrt{\varepsilon_{33}^b/\varepsilon_{11}^b}$ and $\kappa = \sqrt{\varepsilon_{33}\varepsilon_{11}}$, $\kappa_b = \sqrt{\varepsilon_{33}^b\varepsilon_{11}^b}$. After substitution into the boundary conditions we found constants *A, B, C, D*.

(a) For a <u>conductive substrate</u> $\varphi_b(z = h) = 0$ and

$$\varphi_i(\mathbf{r}) = \frac{Q}{2\pi\varepsilon_0} \int_0^\infty dk J_0\left(k\sqrt{x^2+y^2}\right) \frac{\exp\left(-kd - k\frac{z}{\gamma}\right) - \exp\left(-kd - k\frac{2h-z}{\gamma}\right)}{(\varepsilon_e + \kappa) - (\varepsilon_e - \kappa)\exp\left(-\frac{2h}{\gamma}k\right)} \quad \text{(B.5a)}$$

$$\tilde{\varphi}_i(k_1, k_2, z) = \frac{Q}{2\pi\varepsilon_0} \frac{\exp\left(-kd - k\frac{z}{\gamma}\right) - \exp\left(-kd - k\frac{2h-z}{\gamma}\right)}{k\left((\varepsilon_e + \kappa) - (\varepsilon_e - \kappa)\exp\left(-\frac{2h}{\gamma}k\right)\right)} \quad \text{(B.5b)}$$

Here we used that $\dfrac{1}{2\pi}\int_{-\infty}^{\infty} dk_1 \int_{-\infty}^{\infty} dk_2 \exp(-ik_1 x - ik_2 y) f(k) = \int_0^\infty dk\, k\, J_0\left(k\sqrt{x^2+y^2}\right) f(k)$ and substitution $(x, y, z) \to (x_1, x_2, x_3)$.

The Fourier representation $\tilde{E}_j(k_1, k_2, x_3)$ of the electric field $E_j(\mathbf{r}) = -\dfrac{\partial}{\partial x_j}\varphi_i(\mathbf{r})$ acquires the form

$$\tilde{E}_{1,2}(k_1, k_2, x_3) = ik_{1,2} \frac{Q}{2\pi\varepsilon_0} \frac{\exp\left(-kd - k\frac{x_3}{\gamma}\right) - \exp\left(-kd - k\frac{2h-x_3}{\gamma}\right)}{k\left((\varepsilon_e + \kappa) - (\varepsilon_e - \kappa)\exp\left(-\frac{2h}{\gamma}k\right)\right)},$$

$$\tilde{E}_3(k_1, k_2, x_3) = \frac{Q}{2\pi\varepsilon_0} \frac{\exp\left(-kd - k\frac{x_3}{\gamma}\right) + \exp\left(-kd - k\frac{2h-x_3}{\gamma}\right)}{\gamma\left((\varepsilon_e + \kappa) - (\varepsilon_e - \kappa)\exp\left(-\frac{2h}{\gamma}k\right)\right)}.$$

(B.6)

(b) For a <u>dielectric substrate</u>

$$\varphi_i(\mathbf{r}) = \frac{Q}{2\pi\varepsilon_0} \int_0^\infty dk J_0\left(k\sqrt{x_1^2 + x_2^2}\right) \frac{(\kappa_b + \kappa)\exp\left(-kd - k\frac{x_3}{\gamma}\right) - (\kappa_b - \kappa)\exp\left(-kd - k\frac{2h - x_3}{\gamma}\right)}{(\kappa_b + \kappa)(\varepsilon_e + \kappa) - (\kappa_b - \kappa)(\varepsilon_e - \kappa)\exp\left(-\frac{2h}{\gamma}k\right)}$$

(B.7a)

$$\tilde{\varphi}_i(k, x_3) = \frac{Q}{2\pi\varepsilon_0} \frac{(\kappa_b + \kappa)\exp\left(-kd - k\frac{x_3}{\gamma}\right) - (\kappa_b - \kappa)\exp\left(-kd - k\frac{2h - x_3}{\gamma}\right)}{k\left[(\kappa_b + \kappa)(\varepsilon_e + \kappa) - (\kappa_b - \kappa)(\varepsilon_e - \kappa)\exp\left(-\frac{2h}{\gamma}k\right)\right]}$$

(B.7b)

The Fourier representation of the electric field $\tilde{E}_j(k_1, k_2, x_3)$ acquires the form

$$\tilde{E}_{1,2}(k_1, k_2, x_3) = ik_{1,2} \frac{Q}{2\pi\varepsilon_0} \frac{(\kappa_b + \kappa)\exp\left(-kd - k\frac{x_3}{\gamma}\right) - (\kappa_b - \kappa)\exp\left(-kd - k\frac{2h - x_3}{\gamma}\right)}{k\left[(\kappa_b + \kappa)(\varepsilon_e + \kappa) - (\kappa_b - \kappa)(\varepsilon_e - \kappa)\exp\left(-\frac{2h}{\gamma}k\right)\right]},$$

$$\tilde{E}_3(k_1, k_2, x_3) = \frac{Q}{2\pi\varepsilon_0} \frac{(\kappa_b + \kappa)\exp\left(-kd - k\frac{x_3}{\gamma}\right) + (\kappa_b - \kappa)\exp\left(-kd - k\frac{2h - x_3}{\gamma}\right)}{\gamma\left[(\kappa_b + \kappa)(\varepsilon_e + \kappa) - (\kappa_b - \kappa)(\varepsilon_e - \kappa)\exp\left(-\frac{2h}{\gamma}k\right)\right]}.$$

(B.8)

Note, that potential (B.7) transfers into the one given by Eq.(B.5) at $\varepsilon_b \to \infty$ as it should be expected.

Let us calculate the potential on the sample surface below the tip, i.e. $\varphi_i(r = 0)$. After elementary transformations we obtained:

$$\varphi_i(0) = \frac{Q}{2\pi\varepsilon_0(\varepsilon_e + \kappa)} \sum_{m=0}^\infty \left(\frac{\kappa_b - \kappa}{\kappa_b + \kappa}\right)^m \left(\frac{\varepsilon_e - \kappa}{\varepsilon_e + \kappa}\right)^m \left(\frac{\gamma}{\gamma d + 2hm} - \frac{\kappa_b - \kappa}{\kappa_b + \kappa} \frac{\gamma}{\gamma d + 2h(m+1)}\right) \quad (B.9)$$

The asymptotic expansion of Eq.(B.9) has the form:

$$\varphi_i(0) = \begin{cases} \dfrac{Q}{2\pi\varepsilon_0(\varepsilon_e+\kappa)}\left(\dfrac{1}{d}+\dfrac{\kappa}{\varepsilon_e-\kappa}\cdot\dfrac{\gamma}{h}\ln\left(1-\dfrac{\kappa_b-\kappa}{\kappa_b+\kappa}\cdot\dfrac{\varepsilon_e-\kappa}{\varepsilon_e+\kappa}\right)\right), & h \gg \gamma d \\ \dfrac{Q}{2\pi\varepsilon_0(\varepsilon_e+\varepsilon_b)d}, & h \to 0 \\ \dfrac{Q}{2\pi\varepsilon_0(\varepsilon_e+\kappa)d}, & \kappa_b = \kappa \end{cases} \quad (B.10)$$

Pade approximation of Eq.(10) has the form:

$$\varphi_i(0) = \dfrac{Q}{2\pi\varepsilon_0(\varepsilon_e+\kappa)d}\left(1+\left(\dfrac{\kappa_b+\varepsilon_e}{\kappa-\kappa_b}+\dfrac{h}{\gamma d}\dfrac{\varepsilon_e-\kappa}{\kappa}\ln^{-1}\left(1-\dfrac{\kappa_b-\kappa}{\kappa_b+\kappa}\cdot\dfrac{\varepsilon_e-\kappa}{\varepsilon_e+\kappa}\right)\right)^{-1}\right) \quad (B.11)$$

*Effective point charge model*

Under the condition $\varphi_i(0) = U$ ($U$ is potential applied to the tip), the value $Q$ of point charge is the following:

$$Q = Q_\infty\left(\sum_{m=0}^{\infty}\chi^m\left(\dfrac{\gamma d}{\gamma d + 2hm}-\dfrac{\kappa_b-\kappa}{\kappa_b+\kappa}\dfrac{\gamma d}{\gamma d + 2h(m+1)}\right)\right)^{-1}. \quad (B.12)$$

Hereinafter $Q_\infty = 2\pi\varepsilon_0(\varepsilon_e+\kappa)U d$ and $\chi = \left(\dfrac{\kappa_b-\kappa}{\kappa_b+\kappa}\right)\left(\dfrac{\varepsilon_e-\kappa}{\varepsilon_e+\kappa}\right)$. Allowing for potential Pade approximation given by Eq.(B.11) we obtained:

$$Q(h,d) \approx Q_\infty\left(1+\left(\dfrac{\kappa_b+\varepsilon_e}{\kappa-\kappa_b}+\dfrac{h}{\gamma d}\dfrac{\varepsilon_e-\kappa}{\kappa\ln(1-\chi)}\right)^{-1}\right)^{-1}. \quad (B.13)$$

Note that function in paranthesis is $\psi(h,d) \approx 1-\dfrac{\gamma d}{h}\dfrac{\kappa\ln(1-\chi)}{\varepsilon_e-\kappa}$ at $h \gg d$. The latter will be used below for the overall effective charge $Q$ approximation, since $Q(h,d) \approx Q_\infty\left(1-\dfrac{\gamma d}{h}\dfrac{\kappa\ln(1-\chi)}{\varepsilon_e-\kappa}\right)$. Similarly $C(h) \approx C_\infty\left(1-\dfrac{\gamma C_\infty}{2\pi\varepsilon_0 h}\dfrac{\kappa\ln(1-\chi)}{\varepsilon_e^2-\kappa^2}\right)$.

In the effective point charge model the position ($z = -d$) and the value $Q$ of point charge can be found from the conditions that isopotential surface has the curvature $R_0$ in the point $(0,0,0)$ (tip apex touching the sample) and $\varphi_e(0,0,0) = U$ ($U$ is potential applied to the tip). Since $\varphi_{e,x}|_{r=0} = 0$, the curvature can be found as $\varphi_{e,z}/\varphi_{e,xx}|_{r=0}$. After simple integration one can obtain equation, determining the effective distance as

$$R_0 = \frac{V_z(d)}{V_{xx}(d)}, \tag{B.14}$$

where

$$V_z(d) = \kappa \sum_{m=0}^{\infty} \chi^m \left( \left(d + \frac{2h}{\gamma}m\right)^{-2} + \frac{\kappa_b - \kappa}{\kappa_b + \kappa}\left(d + \frac{2h}{\gamma}(m+1)\right)^{-2} \right), \tag{B.15a}$$

$$V_{xx}(d) = \varepsilon_e \sum_{m=0}^{\infty} \chi^m \left( \left(d + \frac{2h}{\gamma}m\right)^{-3} - \frac{\kappa_b - \kappa}{\kappa_b + \kappa}\left(d + \frac{2h}{\gamma}(m+1)\right)^{-3} \right). \tag{B.15b}$$

For the case $h \gg \gamma R_0$ we derived the following approximation

$$d \approx d_\infty \left( 1 - \left(\frac{\gamma R_0}{h}\right)^2 \frac{\varepsilon_e^3}{2(\varepsilon_e - \kappa)\kappa^2} \text{Li}_2(\chi) + \left(\frac{\gamma R_0}{h}\right)^3 \frac{(2\varepsilon_e - \kappa)\varepsilon_e^3}{4(\varepsilon_e - \kappa)\kappa^3} \text{Li}_3(\chi) \right). \tag{B.16}$$

Here $d_\infty = \varepsilon_e R_0/\kappa$, $\text{Li}_n(\chi)$ is the polylogarithmic function $\text{Li}_n(\chi) = \sum_{k=1}^{\infty} \chi^k/k^n$ [37]. Its Pade approximations in the range $|\chi| < 1$ have the form $\text{Li}_2(\chi) \approx \frac{4\chi}{4-\chi}$, $\text{Li}_3(\chi) \approx \frac{8\chi}{8-\chi}$.

Dependence of the effective charge surface separation $d$ on the layer thickness $h$ is shown in Figs.2B for the case of relatively low dielectric permittivity $\kappa = 30$ (a) and high dielectric permittivity $\kappa = 3000$ (b) of surface layer.

**APPENDIX C. Resolution function approach for the case of matched substrate**

In order to find $\tilde{W}_{3jk}(0)$ and $\tilde{W}_{3jk}(qh \gg 1)$ we perform the series expansion of the electric field components given by Eqs.(B.8):

$$\tilde{E}_3(k,\xi_3) = \frac{Q}{2\pi\varepsilon_0} \sum_{m=0}^{\infty} \left(\frac{\kappa_b - \kappa}{\kappa_b + \kappa}\right)^m \left(\frac{\varepsilon_e - \kappa}{\varepsilon_e + \kappa}\right)^m \times$$
$$\times \left( \frac{\exp\left(-kd - \frac{2hk}{\gamma}m - k\frac{\xi_3}{\gamma}\right)}{\gamma(\varepsilon_e + \kappa)} + \frac{\exp\left(-kd - \frac{2hk}{\gamma}(m+1) + k\frac{\xi_3}{\gamma}\right)}{\gamma(\varepsilon_e + \kappa)(\kappa_b + \kappa)/(\kappa_b - \kappa)} \right) \quad \text{(C.1a)}$$

$$\tilde{E}_{1,2}(k_1,k_2,\xi_3) = ik_{1,2}\tilde{\varphi}_i(k,\xi_3) = ik_{1,2}\frac{Q}{2\pi\varepsilon_0} \sum_{m=0}^{\infty} \left(\frac{\kappa_b - \kappa}{\kappa_b + \kappa}\right)^m \left(\frac{\varepsilon_e - \kappa}{\varepsilon_e + \kappa}\right)^m \times$$
$$\times \left( \frac{\exp\left(-kd - \frac{2hk}{\gamma}m - k\frac{\xi_3}{\gamma}\right)}{k(\varepsilon_e + \kappa)} - \frac{\exp\left(-kd - \frac{2hk}{\gamma}(m+1) + k\frac{\xi_3}{\gamma}\right)}{k(\varepsilon_e + \kappa)(\kappa_b + \kappa)/(\kappa_b - \kappa)} \right) \quad \text{(C.1b)}$$

For the case when the piezoelectric layer and substrate (bulk) elastic properties are the same, within the framework of the point charge approach the component $\tilde{W}_{333}(q)$ is:

$$\tilde{W}_{333}(q) = -\int_0^{2\pi} \frac{d\psi}{2\pi} \int_0^{\infty} kdk \int_0^{h} d\xi_3 \exp(-k_q \xi_3)(1 + k_q \xi_3)\tilde{E}_3(k,\xi_3) =$$
$$= -\frac{Q}{2\pi\varepsilon_0} \sum_{m=0}^{\infty} \chi^m \int_0^{2\pi} \frac{d\psi}{2\pi} \int_0^{\infty} kdk \int_0^{h} d\xi_3 \exp(-k_q \xi_3)(1 + k_q \xi_3) \times \quad \text{(C.2)}$$
$$\times \left( \frac{\exp\left(-k\left(d + \frac{2h}{\gamma}m + \frac{\xi_3}{\gamma}\right)\right)}{\gamma(\varepsilon_e + \kappa)} + \frac{\exp\left(-k\left(d + \frac{2h}{\gamma}(m+1) - \frac{\xi_3}{\gamma}\right)\right)}{\gamma(\varepsilon_e + \kappa)(\kappa_b + \kappa)/(\kappa_b - \kappa)} \right)$$

Here $k = \sqrt{k_1^2 + k_2^2}$, $q = \sqrt{q_1^2 + q_2^2}$ and $k_q = \sqrt{k^2 + q^2 - 2kq\cos\psi}$. Performing the integration we obtained for the limiting cases:

$$\widetilde{W}_{333}(0) = \frac{-Q}{2\pi\varepsilon_0(\varepsilon_e + \kappa)} \sum_{m=0}^{\infty} \chi^m \left( \frac{1}{\gamma d + 2hm} + \frac{\kappa_b - \kappa}{\kappa_b + \kappa} \frac{1}{\gamma d + 2h(m+1)} \right) \times$$
$$\times \frac{\gamma h}{(\gamma d + (1 + 2m + \gamma)h)} \left( 1 + \frac{\gamma h}{(\gamma d + (1 + 2m + \gamma)h)} \right) \quad \text{(C.3a)}$$

$$\widetilde{W}_{333}(qd \gg 1) = -\frac{Q}{2\pi\varepsilon_0(\varepsilon_e + \kappa)d} \frac{2}{\gamma q d} + O(\exp(-qh)) \quad \text{(C.3b)}$$

The component $\widetilde{W}_{313}(q)$:

$$\widetilde{W}_{313}(q) = \int_0^{2\pi} \frac{d\psi}{2\pi} \int_0^\infty k dk \int_0^h d\xi_3 \exp(-k_q \xi_3)(-1 - 2\nu + k_q \xi_3)\widetilde{E}_3(k, \xi_3) =$$
$$= \frac{Q}{2\pi\varepsilon_0} \sum_{m=0}^{\infty} \chi^m \int_0^{2\pi} \frac{d\psi}{2\pi} \int_0^\infty k dk \int_0^h d\xi_3 \exp(-k_q \xi_3)(-1 - 2\nu + k_q \xi_3) \times \quad \text{(C.4)}$$
$$\times \left( \frac{\exp\left(-k\left(d + \frac{2h}{\gamma}m + \frac{\xi_3}{\gamma}\right)\right)}{\gamma(\varepsilon_e + \kappa)} + \frac{\exp\left(-k\left(d + \frac{2h}{\gamma}(m+1) - \frac{\xi_3}{\gamma}\right)\right)}{\gamma(\varepsilon_e + \kappa)(\kappa_b + \kappa)/(\kappa_b - \kappa)} \right)$$

Performing the integration we obtained:

$$\widetilde{W}_{313}(0) = \frac{-Q}{2\pi\varepsilon_0(\varepsilon_e + \kappa)} \sum_{m=0}^{\infty} \chi^m \left( \frac{1}{\gamma d + 2hm} + \frac{\kappa_b - \kappa}{\kappa_b + \kappa} \frac{1}{\gamma d + 2h(m+1)} \right) \times$$
$$\times \frac{\gamma h}{(\gamma d + (1 + 2m + \gamma)h)} \left( (1 + 2\nu) - \frac{\gamma h}{(\gamma d + (1 + 2m + \gamma)h)} \right) \quad \text{(C.5a)}$$

$$\widetilde{W}_{313}(qd \gg 1) = \frac{-Q}{2\pi\varepsilon_0(\varepsilon_e + \kappa)d} \frac{2\nu}{\gamma q d} + O(\exp(-qh)) \quad \text{(C.5b)}$$

The component $\widetilde{W}_{351}(q)$:

$$\widetilde{W}_{351}(q) = -\int_0^{2\pi}\frac{d\psi}{2\pi}\int_0^\infty kdk\int_0^h d\xi_3\,\exp(-k_q\xi_3)(k^2 - kq\cos\psi)\xi_3\widetilde{\varphi}_i(k,\xi_3) =$$

$$= \frac{-Q}{2\pi\varepsilon_0}\sum_{m=0}^\infty \chi^m \int_0^{2\pi}\frac{d\psi}{2\pi}\int_0^\infty kdk\int_0^h d\xi_3\,\exp(-k_q\xi_3)(k - q\cos\psi)\xi_3 \times \quad\text{(C.6)}$$

$$\times\left(\frac{\exp\left(-k\left(d+\dfrac{2h}{\gamma}m+\dfrac{\xi_3}{\gamma}\right)\right)}{(\varepsilon_e+\kappa)} - \frac{\exp\left(-k\left(d+\dfrac{2h}{\gamma}(m+1)-\dfrac{\xi_3}{\gamma}\right)\right)}{(\varepsilon_e+\kappa)(\kappa_b+\kappa)/(\kappa_b-\kappa)}\right)$$

Performing the integration we obtained:

$$\widetilde{W}_{351}(0) = \frac{-Q}{2\pi\varepsilon_0(\varepsilon_e+\kappa)}\sum_{m=0}^\infty \chi^m\left(\frac{1}{\gamma d+2hm} - \frac{\kappa_b-\kappa}{\kappa_b+\kappa}\frac{1}{\gamma d+2h(m+1)}\right)\frac{\gamma^3 h^2}{(\gamma d+(1+2m+\gamma)h)^2}$$

(C.7a)

$$\widetilde{W}_{351}(qd \gg 1) = \frac{-Q}{2\pi\varepsilon_0(\varepsilon_e+\kappa)d}\left(\frac{6}{\gamma(qd)^3} - \frac{72}{\gamma^2(qd)^4}\right) + O(\exp(-qh)) \quad\text{(C.7b)}$$

In the case of transversely homogeneous surface layer with constant piezoelectric tensor $d_{ij}^S$, the Fourier image has the form $\widetilde{d}_{ij}^S(q_1,q_2) = 2\pi\delta(q_1)\delta(q_2)d_{ij}^S$. Therefore, their contribution into the vertical displacement is $u_3^S(\mathbf{r}=0) = \dfrac{1}{2\pi}\int_{-\infty}^\infty dq_1\int_{-\infty}^\infty dq_2\,\widetilde{d}_{kj}^S(q_1,q_2)\widetilde{W}_{3jk}^h(-q_1,-q_2) = d_{kj}^S\widetilde{W}_{3jk}(0)$.

Similarly to the case of matched substrate, we obtained from (A.33) the approximate expressions for the response of homogeneous layer:

$$\widetilde{W}_{351}^r(0) \approx \widetilde{W}_{351}^m(0) + \frac{Q}{2\pi\varepsilon_0(\varepsilon_e+\kappa)}\frac{1-2\nu}{2(1-\nu)}\times$$

$$\times\sum_{m=0}^\infty \chi^m\left(\frac{\gamma d}{\gamma d+2h(m+\gamma)} - \frac{\kappa_b-\kappa}{\kappa_b+\kappa}\frac{\gamma d}{\gamma d+2h(m+1+\gamma)}\right)\frac{\gamma^2 h^2}{(\gamma d+(1+2m+3\gamma)h)^2} \quad\text{(C.8)}$$

$$\widetilde{W}_{313}^{r}(0) = \widetilde{W}_{313}^{m}(0) + \frac{(1+\nu)(1-2\nu)}{(1+2\nu)(1-\nu)} \sum_{m=0}^{\infty} \chi^{m} \left( \frac{\gamma d}{\gamma d + 2h(m+\gamma)} + \frac{\kappa_{b} - \kappa}{\kappa_{b} + \kappa} \frac{\gamma d}{\gamma d + 2h(m+1+\gamma)} \right) \times$$
$$\times \left( \frac{h(1+2\nu)}{\gamma d + (1+2m+3\gamma)h} + \frac{\gamma h^{2}}{(\gamma d + (1+2m+3\gamma)h)^{2}} \right) \frac{Q}{2\pi\varepsilon_{0}(\varepsilon_{e} + \kappa)} \quad (C.9)$$